# Condensed-matter equation of states covering a wide region of pressure studied experimentally

Elijah E. Gordon[1], Jürgen Köhler[2] & Myung-Hwan Whangbo[1,*]

[1] Department of Chemistry, North Carolina State University, Raleigh, NC 27695-8204, USA

[2] Max-Planck-Institut für Festkörperforschung, D-70569 Stuttgart, Germany

**Abstract**

**The relationships among the pressure $P$, volume $V$, and temperature $T$ of solid-state materials are described by their equations of state (EOSs), which are often derived from the consideration of the finite-strain energy or the interatomic potential.[1-3] These EOSs consist of typically three parameters to determine from experimental $P$-$V$-$T$ data by fitting analyses. In the empirical approach to EOSs, one either refines such fitting parameters or improves the mathematical functions[3-5] to better simulate the experimental data. Despite over seven decades of studies on EOSs, none has been found to be accurate for all types of solids over the whole temperature and pressure ranges studied experimentally.[3,6,7] Here we show that the simple empirical EOS, $P = \alpha_1(PV) + \alpha_2(PV)^2 + \alpha_3(PV)^3$, in which the pressure $P$ is indirectly related to the volume $V$ through a cubic polynomial of the energy term $PV$ with three fitting parameters $\alpha_1 - \alpha_3$, provides accurate descriptions for the $P$-vs-$V$ data of condensed matter in a wide region of pressure studied experimentally even in the presence of phase transitions.**

# Introduction

One of the most important issues in condensed matter sciences, particularly, in geology and geophysics, is to accurately predict the structural and physical properties of solids under high pressure and temperature. In general, a solid-state material under high pressure and temperature can exhibit properties quite different from those found at ambient conditions. At a given temperature $T$, a solid under external pressure $P$ decreases its volume $V$ with increasing $P$, but $V$ changes a lot more slowly than does $P$. The pressure-induced volume decrease may require a change in the structure type (i.e., the pattern of the relative atom arrangements in a repeat unit cell) thereby causing a structural phase transition and an associated physical property change. For example, when $P$ is increased at room temperature, elemental chalcogen Te,[8-14] Se,[15-20] or S [15,21-25] undergoes a number of structural phase transitions while its electrical property changes from insulating at ambient pressure to metallic and superconducting at high pressure.[26,27] Hydrogen sulfide $H_2S$ is a diamagnetic molecular species at ambient conditions, but is converted, under the pressure of over ~110 GPa, to a condensed phase that becomes superconducting at ~200 K,[25,28] the highest among all superconductors known so far. An isothermal EOS relates $P$ and $V$ at a certain temperature $T$. Over the past 70 years the $P$-vs-$V$ data have been studied for a variety of solids in various pressure ranges (e.g., see **Table 1**), and their EOSs have been examined. So far, however, no isothermal EOS is applicable to all types of solids and is accurate over the whole range of pressure studied especially when a solid undergoes several structural phase transitions in the pressure region studied.

With increasing pressure $P$, the volume $V$ of a solid under pressure changes very slowly compared with the pressure change. The shortcomings of the known EOSs originate essentially from the attempts to relate the fast changing variable $P$ to a very slowly changing variable $V$. These problems can be circumvented if the pressure change is related to a pressure-induced energy change that is associated with the volume $V$ and also varies nearly at the same rate as does $P$. The energy term $PV$ satisfies these two requirements because, while increasing $P$, the volume $V$ of a solid under pressure $P$ decreases very slowly so that the term $PV$ changes nearly as fast as $P$ in the entire range of $P$. Furthermore, at a given $P$, the term $PV$ is determined by the value of $V$, not by how the atoms are arranged within the volume so that the term $PV$ cannot be overly sensitive to phase transitions. With increasing $P$, the term $PV$ should increase slightly more slowly than does $P$ because $V$ decreases slightly under pressure. Therefore, it should be possible to accurately describe the $P$-vs-$V$ data of any solid over the entire pressure range studied experimentally by the simple EOS,

$$P = \alpha_1(PV) + \alpha_2(PV)^2 + \alpha_3(PV)^3 + \cdots, \qquad (1)$$

which expands $P$ as a polynomial of $PV$, where the constants $\alpha_i$ (i = 1, 2, 3, etc.) are the fitting parameters. For those familiar with the traditional EOSs, use of Eq. 1 is quite unconventional because $P$ is expanded in terms of the variable $PV$ that contains itself. However, our goal is to find an accurate, though indirect, relationship between $P$ and $V$ valid for the entire pressure region studied experimentally by way of determining an accurate relationship between $P$ and $PV$. From the resulting $P$-vs-$V$ relationship, one can derive an accurate description of other thermodynamic quantity such as bulk modulus, as we demonstrated in our work. When this expression is truncated to $P = \alpha_1(PV) + \alpha_2(PV)^2$, $P = \alpha_1(PV) + \alpha_2(PV)^2 + \alpha_3(PV)^3$, and $P = \alpha_1(PV) + \alpha_2(PV)^2 + \alpha_3(PV)^3 + \alpha_4(PV)^4$, we obtain the quadratic, cubic and quartic approximations for Eq. 1. Here we establish that the cubic approximation with three fitting parameters $\alpha_1 - \alpha_3$ is accurate enough in describing the $P$-vs-$V$ data for condensed matter in a wide region of pressure experimentally probed even in the presence of several structural phase transitions.

## Results

**Formulation of the EOS.** In testing whether an isothermal EOS is accurate over the entire range of pressure studied experimentally, the ideal systems to analyze would be elemental chalcogens Te, Se and S because they have been studied at room temperature over wide pressure ranges (i.e., 0 – 330 GPa for Te, 0 – 150 GPa for Se, and 0 – 213 GPa for S) (see **Table 1**), because each chalcogen undergoes a number of phase transitions with increasing $P$, and because their room-temperature atomic structures are known under widely different pressures. For each chalcogen, we begin our analysis by first determining the relative energies of its known atomic structures at various $P$ on the basis of density functional theory (DFT) calculations, the details of which are described in Methods. We summarize the space groups of the known atomic structures at various pressures $P$ (mostly around the room temperature), the volumes $V$ per atom, the energies $PV$ per atom, and the calculated electronic energies $E$ per atom in Section 1(a)-(c) of the supporting information (SI). The calculated electronic structures are also presented in terms of density of states (DOS) plots in Section 1(d)-(f) of the SI. As anticipated, with increasing pressure, the DOS plot for each chalcogen is shifted toward the higher energy while the band gap present at low pressure disappears at high pressure.

The calculated energy $E$ for Te is plotted as a function of $P$ in **Fig. 1a** (those for Se and S in Section 2(a), (b) of the SI), which shows a reasonable linear relationship, $E \approx a_1 P + a_0$, with slope $a_1$ and intercept $a_0$. **Fig. 1a** also plots the enthalpy $H = E + PV$ per atom as a function of $P$ using $a_0$ as the intercept. This plot also exhibits a reasonable linear relationship, $H \approx b_1 P + a_0$, with the slope $b_1 \gg a_1$ for all chalcogens. At a given temperature, therefore, the energy term $PV = H - E$ varies almost linearly with $P$, i.e., $P \approx PV/(b_1 - a_1)$. Unlike the case of gaseous substances for which the *PV* term is a constant independent of *P* at a given *T*, the *PV* term for each condensed-phase chalcogen increases almost linearly with *P* because, compared with the rate of change in *P*, that in *V* is very small. Nevertheless, the *H*-vs-*P* plot for each chalcogen is slightly concave down with respect to the base line, $b_1P$. As already pointed out, this reflects that, with increasing *P*, the rate of change in *P* is slightly greater than that in *PV* due to a pressure-induced decrease in *V*. The latter allows one to expand *P* as a power series of *PV* as expressed in Eq. 1. Indeed, the *P*-vs-*V* data points used for constructing the *H*-vs-*P* plots are very well described, for example, by the quadratic approximation of Eq. 1 (see Section 2(c) of the SI). As will be discussed below, the nonlinear terms of Eq. 1, e.g., $\alpha_2(PV)^2$ and $\alpha_3(PV)^3$, are related to how the volume *V* of a solid decreases under pressure *P*.

To test the applicability of the isothermal EOS, Eq. 1, in a wide region of pressure studied experimentally, we first analyze the experimental *P*-vs-*V* data available in the literature for each chalcogen. The *P*-vs-*PV* plot for Te, presented in **Fig. 1b**, reveals that the experimental points in the 0 – 330 GPa region [8-14] are very well described by the cubic approximation. The fitting curves from the quadratic and quartic approximations are not shown because they are practically impossible to distinguish, with naked eye alone, from that of the cubic approximation. The same conclusion is reached for Se [15-20] and S [15,21-25] (see Section 2(a), (b) of the SI for Se and S, respectively). The fitting coefficients $\alpha_1 - \alpha_3$ obtained for Te, Se and S resulting from the cubic approximation are summarized in **Table 2**, and those from the quadratic, cubic and quartic approximations are compared in Section 2(d) of the SI. The coefficients $\alpha_1$ and $\alpha_2$ are always positive, and $\alpha_1 \gg \alpha_2 \gg |\alpha_3|$ with $\alpha_2/\alpha_1 \approx 10^{-3}$ and $|\alpha_3|/|\alpha_2| \approx 10^{-4}$.

**Error analysis.** To assess the accuracies of the quadratic, cubic and quartic approximations for Eq. 1, we analyze the pressure-dependence of the absolute errors, $\Delta P = P_{\text{calc}} - P_{\text{expt}}$, as well as that of the % errors, $100 \times \Delta P/P_{expt}$, where $P_{expt}$ is the pressure observed experimentally, and $P_{calc}$ the one calculated from the fitting equations. The pressure-dependence of the % errors for Te is shown in **Fig. 1c**, and those for Se and S in Section 2(a), (b) of the SI. The maximum % error is smaller than 5.5 % in the 10.9 – 330 GPa region for the cubic and quartic approximations, but smaller than 5.4 % in the 38 – 330 GPa region for the quadratic approximation. The % errors are large in the low $P$ region, but it should be pointed out that the associated absolute errors are rather small (for example, for $P_{expt}$ = 0.98 GPa, the corresponding $P_{calc}$ values are 1.44, 1.22 and 1.22 GPa from the quadratic, cubic and quartic approximations, respectively) (see Section 2(e) of the SI). In general, the cubic and quartic approximations are similar in accuracy, and are more accurate than the quadratic approximation especially in the low $P$ region. A similar conclusion is reached from the % error and absolute error plots calculated for Se in the 0 – 140 GPa range and for S in the 0 – 213 GPa range (see Section 2(a), (b) of the SI).

**Applicability to other condensed matter.** The above analyses of the experimental $P$-vs-$V$ data for elemental chalcogens suggest strongly that the cubic approximation of the isothermal EOS, Eq. 1, can accurately describe the experimental $P$-vs-$V$ data for various solids in the whole pressure range studied experimentally. To establish this point, we examine the experimental $P$-vs-$V$ data for various solid-state condensed matter listed in **Table 1**, which include the elemental Sn, the transition-metals Au and Cu, the alkali halides LiF, NaF, NaCl and CsCl, ice VII, the oxides MgO and $MgSiO_3$, the noble gases Ar, Kr and Xe, as well as molecular hydrogen $H_2$. As representative examples of these analyses, we discuss the oxides MgO[29-35] and $MgSiO_3$.[36-41] The isothermal $P$-vs-$V$ relationships of these oxides have been extensively studied because they are the end members of (Mg,Fe)O[42,43] and $(Mg,Fe)SiO_3$ perovskite,[42,44] which are the important components of the Earth's lower mantle. The $P$-vs-$V$ relationships for MgO [29-35] were examined at room temperature in the 0 – 140 GPa range, and those for $MgSiO_3$ [36-41] at room temperature in the 0 – 300 GPa range. The *P*-vs-*PV* plots and the % error vs. *P* plots for MgO and $MgSiO_3$ are presented in **Fig. 2**. The maximum % error is smaller than ~0.6 % for MgO, and smaller than ~1 % for $MgSiO_3$, in the entire pressure ranges studied experimentally. For the remainder of the solids listed in **Table 1**, our results are summarized in Section 3(a)-(k) of the SI. The fitting coefficients $\alpha_1 - \alpha_3$ obtained for the solid-state condensed matter of **Table 1** from the cubic approximation are listed in **Table 2** together with the maximum % errors. It is clear that the cubic approximation of the isothermal EOS, Eq. 1, provides an accurate description in the entire pressure regions examined experimentally. (Hereafter, the cubic approximation of Eq. 1 will be used without further mentioning.)

The isothermal EOS, Eq. 1, is also applicable to non-solid-state condensed matter. As examples, we analyze the experimental *P*-vs-*V* data for the polymer, poly(ε-caprolactone) (PCL),[45] determined at 100.6 °C in the 0 – 0.2 GPa region as well as those for liquid $H_2O$ [46] determined at 15 °C, 25 °C and 35 °C in the 0 – 0.1 GPa region. Our results summarized in Section 3(m)-(p) of the SI show that the maximum % error for the polymer is smaller than 0.3 %, and that for liquid water is practically zero, in the entire pressure region studied. Clearly, Eq. 1, provides an accurate description of the *P*-vs-*V* relationship for these materials. It should be pointed out that the ideal gas law is a special case of Eq. 1, when the PV term is a constant independent of *P*.

**Discussion**

**Bulk modulus.** Now that Eq. 1 provides an isothermal EOS accurate for the entire pressure range studied for a given system, we search for a simple expression for the corresponding bulk modulus $B$ valid for the entire pressure region. In the cubic approximation, Eq. 1 is a quadratic equation of $P$, from which $P$ is written in terms of $V$ as

$$P = -\left(\frac{\alpha_2}{2\alpha_3}\right)\left(\frac{1}{V}\right) + \left(\frac{\sqrt{(\alpha_2^2 - 4\alpha_1\alpha_3)V^2 + 4\alpha_3 V}}{2\alpha_3 V^2}\right). \tag{2}$$

Thus, the bulk modulus $B$ is expressed as

$$\begin{aligned}B = -V\left(\frac{\partial P}{\partial V}\right)_T &= -\left(\frac{\alpha_2}{2\alpha_3 V}\right) - \left(\frac{1}{2\alpha_3 V}\right)\left(\frac{(\alpha_2^2 - 4\alpha_1\alpha_3)V + 2\alpha_3}{\sqrt{(\alpha_2^2 - 4\alpha_1\alpha_3)V^2 + 4\alpha_3 V}}\right) \\ &+ \left(\sqrt{(\alpha_2^2 - 4\alpha_1\alpha_3)V^2 + 4\alpha_3 V}\right)\left(\frac{1}{\alpha_3 V^2}\right).\end{aligned} \tag{3}$$

This equation expresses the bulk modulus as a function of volume, namely, $B(V)$. For each value of $V$, however, there is a unique value of $P$ associated with it, so that the $B(V)$ vs. $V$ relationship can be easily converted to the corresponding $B(P)$ vs. $P$ relationship. For convenient use of this relationship, we fit the $B(P)$-vs.-$P$ relationship by the polynomial,

$$B(P) = B_0 + B_1 P + B_2 P^2 + B_3 P^3. \tag{4}$$

The $B(P)$-vs-$P$ plot thus-obtained for Te is presented in **Fig. 1d**. For other solids listed in **Table 1**, the $B(P)$-vs-$P$ plots are presented in Section 3(a)-(r) of the SI. The coefficients $B_0$, $B_1$, $B_2$ and $B_3$ determined for each condensed matter are summarized in **Table 3**, which also lists the calculated bulk modulus at $P = 0$, referred to as $B_{0,calc}$, for each system. The $B_0$ deviates from the $B_{0,calc}$ because the polynomial fitting (Eq. 4) poorly describe the low-pressure region. Nevertheless, the $B_0$ and $B_{0,calc}$ values are quite similar for all systems except for Te and Se.

For every system, our analysis leads to only one $B_0$ value because the entire pressure region studied is represented by the single fitting curve, Eq. 4. In the traditional study for a system undergoing several phase transitions, each phase covering a certain pressure region (say, $P_1$ to $P_2$) is described by the EOS covering only the pressure region $P_1 - P_2$. The resulting EOS for each different phase generates the bulk modulus, which we will refer to as $B_{0,\text{expt}}$, and the virtual volume $V_0$. It is the $B_{0,\text{expt}}$ obtained for the "first" phase (i.e., the phase stable in the lowest pressure region for which $P_1 = 0$) that should be compared with the $B_0$ or the $B_{0,\text{calc}}$ value obtained from our EOS. For systems with several phases, **Table 3** lists only the $B_{0,\text{expt}}$ values of their first phases. Clearly, these $B_{0,\text{expt}}$ values are well described by the $B_0$ and/or $B_{0,\text{calc}}$ values determined from our EOS analyses. The $B_{0,\text{expt}}$ values found for the various phases of Te, Se and S can be accounted for in terms of our EOS analyses as presented in Section 4 of the SI.

**Qualitative meaning of the EOS.** To gain insight into meaning of the EOS, Eq. 1, we rewrite it in a slightly different form. In general, $\alpha_1(PV) \gg \alpha_2(PV)^2 \gg \alpha_3(PV)^3$ so that $P \approx \alpha_1(PV)$. By using this approximation, Eq. 1 is rewritten as

$$V \approx \frac{1}{\alpha_1}\left(1 - \frac{\alpha_2}{\alpha_1^2}P - \frac{\alpha_3}{\alpha_1^3}P^2\right) = V_c\left(1 - \alpha_2 V_c^2 P - \alpha_3 V_c^3 P^2\right) \tag{5}$$

where the volume $V_c$ is defined as $V_c \equiv 1/\alpha_1$, which is very close to the volume at zero-pressure, $V_0$. Eq. 5 reveals that the decrease in the volume of a condensed matter under pressure is a polynomial function of $P$. Since $\alpha_2(PV_c)^2 \gg |\alpha_3|(PV_c)^3$, the term $\alpha_2 V_c^2 P$ dominates over $\alpha_3 V_c^3 P^2$. Namely, the volume decreases with increasing $P$. The $\alpha_3 V_c^3 P^2$ term compensates the overcorrection (when $\alpha_3 < 0$), or the under-correction (when $\alpha_3 > 0$), given by the $\alpha_2 V_c^2 P$ term. In essence, the EOS, Eq. 1, reveals that the volume of a solid under pressure can be described as a polynomial function of pressure.

**Summary**

We presented the empirical equation of state that accurately describes the pressure-versus-volume data for various types of condensed matter in a wide region of pressure studied experimentally. This is also true for systems undergoing several phase transitions in the pressure region studied. This equation of states results from the fact that the pressure change of a condensed matter is accurately described by a cubic polynomial of the pressure times the volume.

**Methods**

Our non-spin-polarized DFT calculations employed the frozen-core projector augmented wave method[47,48] encoded in the Vienna ab initio simulation package,[49] and the generalized-gradient approximation of Perdew, Burke and Ernzerhof[50] for the exchange-correlation functional. To ensure the accuracies of the calculations, a high plane-wave cut-off energy of 1000 eV was used, and the Brillouin zone associated with each repeat unit cell was sampled by a large number of k-points. For example, the $R\bar{3}m$ structure of S at 206 GPa was calculated by using a set of 24×24×24 k-points. Our calculations for Te, Se and S employed their reported crystal structures under various pressures, except for the 160 and 173 GPa structures of S as described in Section 1(c). The threshold for the self-consistent-field energy convergence was set at $10^{-8}$ eV for all structures of Se, Te and S. For the optimization of the structures of S at 160 and 173 GPa, the threshold for the force convergence at each atom was set at 0.005 eV/Å.

**Acknowledgments**

This research used resources of the National Energy Research Scientific Computing Center, a DOE Office of Science User Facility supported by the Office of Science of the U.S. Department of Energy under Contract No. DE-AC02-05CH11231. MHW thanks Prof. Jerry L. Whitten for invaluable discussion concerning the qualitative meaning of the isothermal EOS, Eq. 2.

**Author contributions**

This work was conceived by M.-H.W. and J.K., and E.E.G. carried out all DFT calculations. The empirical EOS was realized by M.-H.W. and E.E.G. while analyzing the computational results. Further testing of the EOS was designed by M.-H.W. and J.K., and E.E.G. carried out the analyses. All authors participated in the discussion of the draft written by M.-H.W.

**Competing financial interests**

The authors declare that they have no competing financial interests.

**Correspondence**

Correspondence and requests for materials should be addressed to M.-H.W. (email: mike_whangbo @ncsu.edu).

Table 1. The pressure ranges (in GPa) and temperature (K) employed to examine the isothermal $P$-vs-$V$ relationships for various condensed matter.[†,‡]

| System | Pressure range (GPa) | Temperature (K) |
|---|---|---|
| Te | 0,[11] 0 – 4,[10] 4.5,[12] 8,[13] 4 – 36,[9] 33,[14] 30 – 330[8] | 298 |
| Se | 0,[17] 0 – 10,[15] 4.6,[18] 23,[19] 28,[19] 87.9,[20] 140,[19] 5 – 150[16] | 293 – 298 |
| S | 0,[22] 0 – 30,[15] 35 – 87,[15] 89.4,[23] 145,[21] 160,[24] 173,[25] 206.5,[21] 88 – 213[21] | 293 – 298 |
| Sn | 0 – 120[a-c] | 298 |
| Au | 4 – 70[d] | 298 |
| Cu | 7 – 95[e] | 293 |
| LiF | 0 – 4,[h] 1 – 9,[f] 0 – 30[g] | 298 |
| NaF | 1 – 9,[f] 0 – 38[g] | 298 |
| NaCl | 0 – 4[i] | 298 |
| CsCl | 0 – 5,[k] 1 – 9,[f] 0 – 45[j] | 298, 293 |
| Ice VII | 3 – 19,[l] 4 – 128[m] | 300 |
| MgO | 0 – 8,[33] 0 – 11,[30] 0 – 20,[32] 0 – 24,[31] 0 – 52,[29] 4 – 120,[35] 10 – 140[34] | 298 |
| MgSiO$_3$ | 0 – 10,[40] 0 – 20,[36] 0 – 55,[37] 36 – 83,[38] 29 – 91,[39] 100 – 300[41] | 298 |
| Ar | 0 – 2[6,n] | 40 |
| Kr | 0 – 2[6,n] | 60 |
| Xe | 0 – 2[6,n] | 60 |
| H$_2$ | 0 – 2.6[47] | 4.2 |
| D$_2$ | 0 – 2.6[47] | 4.2 |
| PCL | 0 – 0.2[45] | 373.6 |
| Liquid H$_2$O | 0 – 0.1[46] | 288 |
| Liquid H$_2$O | 0 – 0.1[46] | 298 |
| Liquid H$_2$O | 0 – 0.1[46] | 308 |

[†] In our analysis for Sn, the α-Sn phase was excluded because it exists below 286 K, but all other phases of Sn that exist at room temperature are included.

[‡] For the references a – p, see Section 3 of the SI.

Table 2. The coefficients $\alpha_1 - \alpha_3$ of the isothermal EOS, $P = \alpha_1(PV) + \alpha_2(PV)^2 + \alpha_3(PV)^3$, obtained for various condensed matter. The range of the pressure $P$ used for each fitting analysis and the maximum % error found in the pressure range are also given.

(a) Solid-state condensed matter ($P$ in GPa, and $V$ in Å$^3$)

|  | $10^2 \times \alpha_1$ | $10^5 \times \alpha_2$ | $10^9 \times \alpha_3$ | $P$ range | Max. % error [b] |
|---|---|---|---|---|---|
| Te | 3.785 | 1.501 | -1.130 | 0 – 330 | 5.5 ($P > 10.9$)<br>29 ($P < 10.9$) |
| Se | 4.782 | 2.421 | +1.325 | 0 – 150 | 3.6 ($P > 23$)<br>44 ($P < 23$) |
| S[a] | 3.587 | 7.987 | -16.20 | 0 – 213 | 4.6 ($P > 36$)<br>13 ($P < 36$) |
| Sn | 3.978 | 1.897 | -2.039 | 0 – 120 | 2.2 ($P > 10.3$)<br>8.0 ($P < 10.3$) |
| Au | 5.839 | 2.079 | -4.474 | 4 – 70 | 1.9 |
| Cu | 8.543 | 4.582 | -10.02 | 7 – 95 | 1.0 ($P > 56$)<br>6.5 ($P < 56$) |
| LiF | 6.064 | 6.646 | -59.75 | 0 – 30 | 0.98 |
| NaF | 4.068 | 2.922 | -13.88 | 0 – 38 | 1.5 |
| NaCl | 2.232 | 2.006 | -11.35 | 0 – 4 | 0.1 |
| CsCl | 1.518 | 0.6655 | -0.8938 | 0 – 45 | 6.4 |
| Ice VII | 4.898 | 8.732 | -24.99 | 3 – 128 | 3.0 |
| MgO | 5.379 | 1.666 | -1.192 | 0 – 142 | 0.6 |
| MgSiO$_3$ | 2.437 | 0.2434 | -0.04751 | 0 – 265 | 1.2 |
| Ar | 2.682 | 18.55 | -876.3 | 0 – 2 | 2.5 |
| Kr | 2.198 | 12.04 | -459.7 | 0 – 2 | 2.2 |
| Xe | 1.728 | 6.493 | -175.2 | 0 – 2 | 1.6 |
| H$_2$ | 3.323 | 104.0 | -7594 | 0 – 2.6 | 3.0 ($P > 0.34$)<br>26 ($P < 0.34$) |
| D$_2$ | 3.552 | 99.32 | -7221 | 0 – 2.6 | 2.0 ($P > 0.15$)<br>18 ($P < 0.15$) |

(b) Non-solid-state condensed matter ($P$ in bar, and $V$ in Å$^3$)

|  | $10^2 \times \alpha_1$ | $10^5 \times \alpha_2$ | $10^9 \times \alpha_3$ | $P$ range | Max. % error [b] |
|---|---|---|---|---|---|
| PCL | 1.03976 | 6.290 | -7.651 | 0 – 2000 | 0.3 |
| H$_2$O, 15 °C | 0.03343 | 5.222×10$^{-3}$ | -1.073×10$^{-4}$ | 0 – 1000 | 0 |
| H$_2$O, 25 °C | 0.03336 | 5.026×10$^{-3}$ | -9.835×10$^{-5}$ | 0 – 1000 | 0 |
| H$_2$O, 35 °C | 0.03326 | 4.901×10$^{-3}$ | -9.451×10$^{-5}$ | 0 – 1000 | 0 |

[a] The $P$-vs-$V$ data points of the metastable S-III phase between 3 – 58 GPa (obtained by quenching) reported in ref. 15 were not included in the $P$-vs-$PV$ plot. However, including them does not change the quality of the fitting analysis.

[b] Unless mentioned otherwise, the maximum % error refers to the entire pressure region studied.

Table 3. The coefficients of the bulk modulus formulas, $B(P) = B_0 + B_1P + B_2P^2 + B_3P^3$, obtained for various condensed matter using $P$ and $B(P)$ in GPa units. For comparison, the $B_{0,\text{expt}}$ and $B_{0,\text{calc}}$ values are also listed (see the text for the definition). [a,b]

|   | $B_{0,\text{expt}}$ | $B_{0,\text{calc}}$ | $B_0$ | $B_1$ | $B_2$ | $B_3$ |
|---|---|---|---|---|---|---|
| Te | 24 (P = 2)[14] | 37.68 | 62.48 | 4.135 | -7.200×10$^{-3}$ | 2.278×10$^{-5}$ |
| Se | 48.1 (P = 7.7)[18] | 28.09 | 46.99 | 5.311 | -4.021×10$^{-2}$ | 1.516×10$^{-4}$ |
| S | 14.5[21] | 20.10 | 25.28 | 2.575 | -2.635×10$^{-4}$ | 2.837×10$^{-5}$ |
| Sn | 54.6,[o] 55.4,[o] 54.92[p] | 64.31 | 59.58 | 5.262 | -4.100×10$^{-2}$ | 2.461×10$^{-4}$ |
| Au | 166.65[6,d] | 192.09 (P = 4.42) | 171.09 | 3.516 | 2.146×10$^{-2}$ | -2.613×10$^{-5}$ |
| Cu | 133[6,q] | 184.48 (P = 7.2) | 156.32 | 4.025 | -4.760×10$^{-3}$ | 6.803×10$^{-5}$ |
| LiF | 66.4[6,q] | 57.94 | 58.06 | 4.136 | 1.520×10$^{-2}$ | 4.610×10$^{-3}$ |
| NaF | 46.1[6,q] | 53.20 | 52.44 | 5.685 | -0.1005 | 3.900×10$^{-3}$ |
| NaCl | 23.5[6] | 25.20 (P = 0.106) | 24.76 | 4.440 | -8.684×10$^{-2}$ | 2.592×10$^{-2}$ |
| CsCl | 16.8[6,q] | 27.57 | 28.79 | 4.896 | -7.036×10$^{-2}$ | 1.290×10$^{-3}$ |
| Ice VII | 23.9[6,q] | 38.71 (P = 3.16) | 30.37 | 2.980 | -1.950×10$^{-3}$ | 7.809×10$^{-5}$ |
| MgO | 153 – 182[30] | 170.66 | 171.63 | 3.562 | -4.270×10$^{-3}$ | 2.269×10$^{-5}$ |
| MgSiO$_3$ | 200 – 340[41] | 253.84 | 255.85 | 2.995 | 1.370×10$^{-3}$ | 2.150×10$^{-6}$ |
| Ar | 2.35[6,n] | 3.49 | 3.482 | 6.053 | -1.732 | 0.9499 |
| Kr | 2.49[6,n] | 3.66 | 3.658 | 5.909 | -1.514 | 0.8780 |
| Xe | 3.02[6] | 4.52 (P = 0.05) | 4.218 | 5.942 | -1.373 | 0.7384 |
| H$_2$ | 0.170[51], 0.174[52] | 0.44 | 0.5707 | 5.002 | -1.521 | 0.5286 |
| D$_2$ | 0.315[51], 0.337[52] | 0.70 | 0.7980 | 4.979 | -1.427 | 0.4953 |
| PCL | 3.01[45] | 1.683 | 1.679 | 8.491 | -7.773 | 86.34 |
| H$_2$O (15 °C) | 2.140[46] | 2.138 | 2.138 | 5.627 | 2.734 | 4.710 |
| H$_2$O (25 °C) | 2.210[46] | 2.213 | 2.213 | 5.605 | 2.388 | 5.164 |
| H$_2$O (35 °C) | 2.250[46] | 2.256 | 2.256 | 5.632 | 2.277 | 5.649 |

[a] For the references d, l, and n – q, see Section 3 of the SI.

[b] Unless otherwise stated, the $B_{0,\text{expt}}$ and $B_{0,\text{calc}}$ refer to the values at $P = 0$. When these values are obtained at a nonzero $P$, the value of $P$ is specified in the parenthesis.

**Figure captions**

Figure 1. (a) The $E$-vs-$P$ and $H$-vs-$P$ plots calculated for Te, where $E$, $H$ and $PV$ are in eV units. The fitting coefficients $a_1$, $a_0$ and $b_1$ for the linear plots $E = a_1P + a_0$ and $H = b_1P + a_0$ are respectively -3.0238, 0.0246 and 0.1589. (b) The $P$-vs-$PV$ plot constructed from the experimental $P$-vs-$V$ data for Te, where the solid line is the fitting curve obtained by using the cubic approximation for the EOS, Eq. 1. (c) The pressure-dependence of the % error, $100\times(P_{calc} - P_{expt})/P_{expt}$, of the $P$-vs-$PV$ plot obtained for Te by using the cubic approximation of the EOS, Eq. 1. (d) The pressure dependence of the bulk modulus $B(P)$ calculated for Te.

Figure 2. (a, c) The $P$-vs-$PV$ plots constructed from the experimental $P$-vs-$V$ data for MgO and MgSiO$_3$, respectively, where the solid lines are the fitting curves obtained by using the cubic approximation of the EOS, Eq. 1. (b, d) The plots of the % errors vs. pressure obtained for MgO and MgSiO$_3$, respectively, by using the cubic approximation of the EOS, Eq. 1.

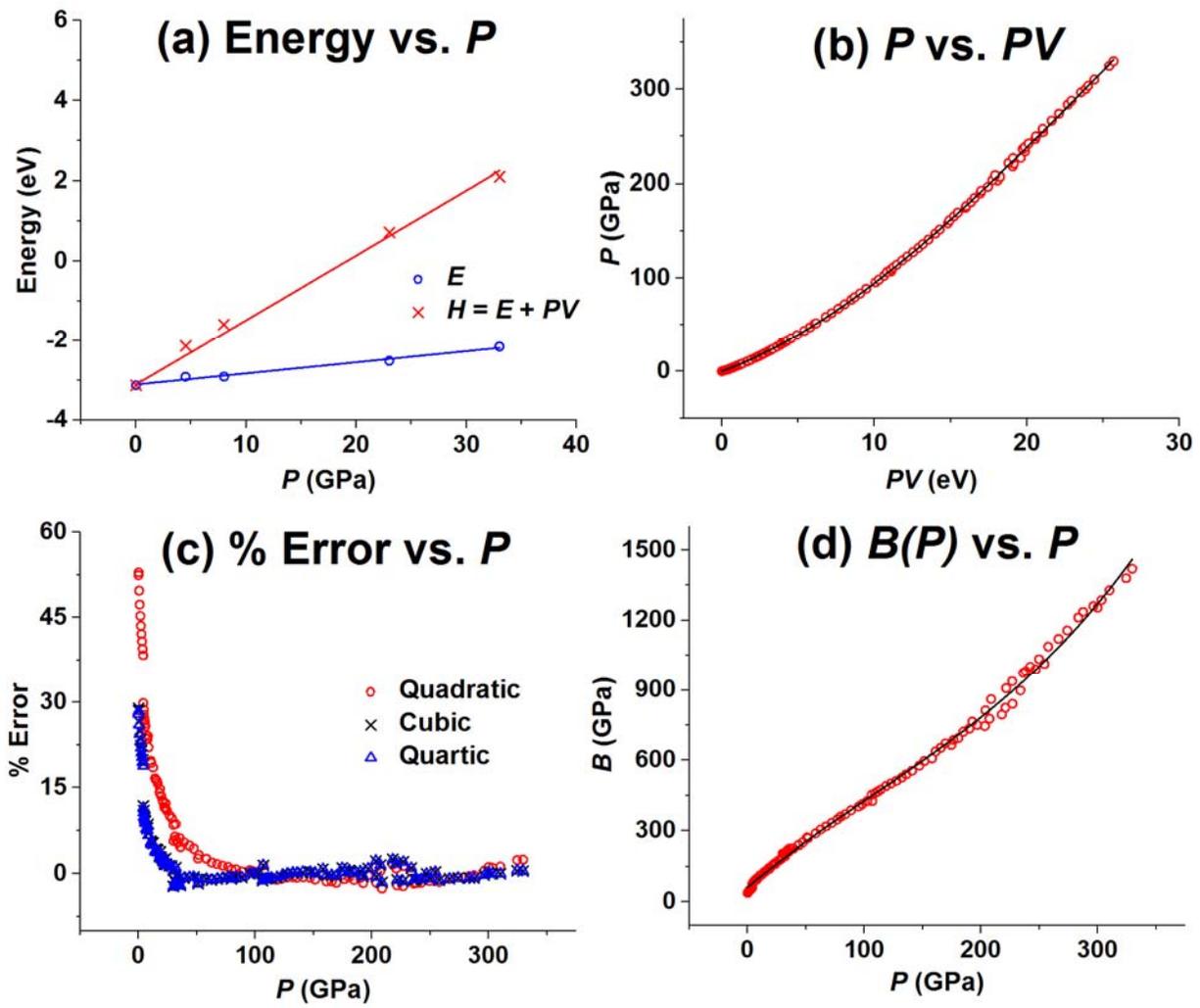

Figure 1

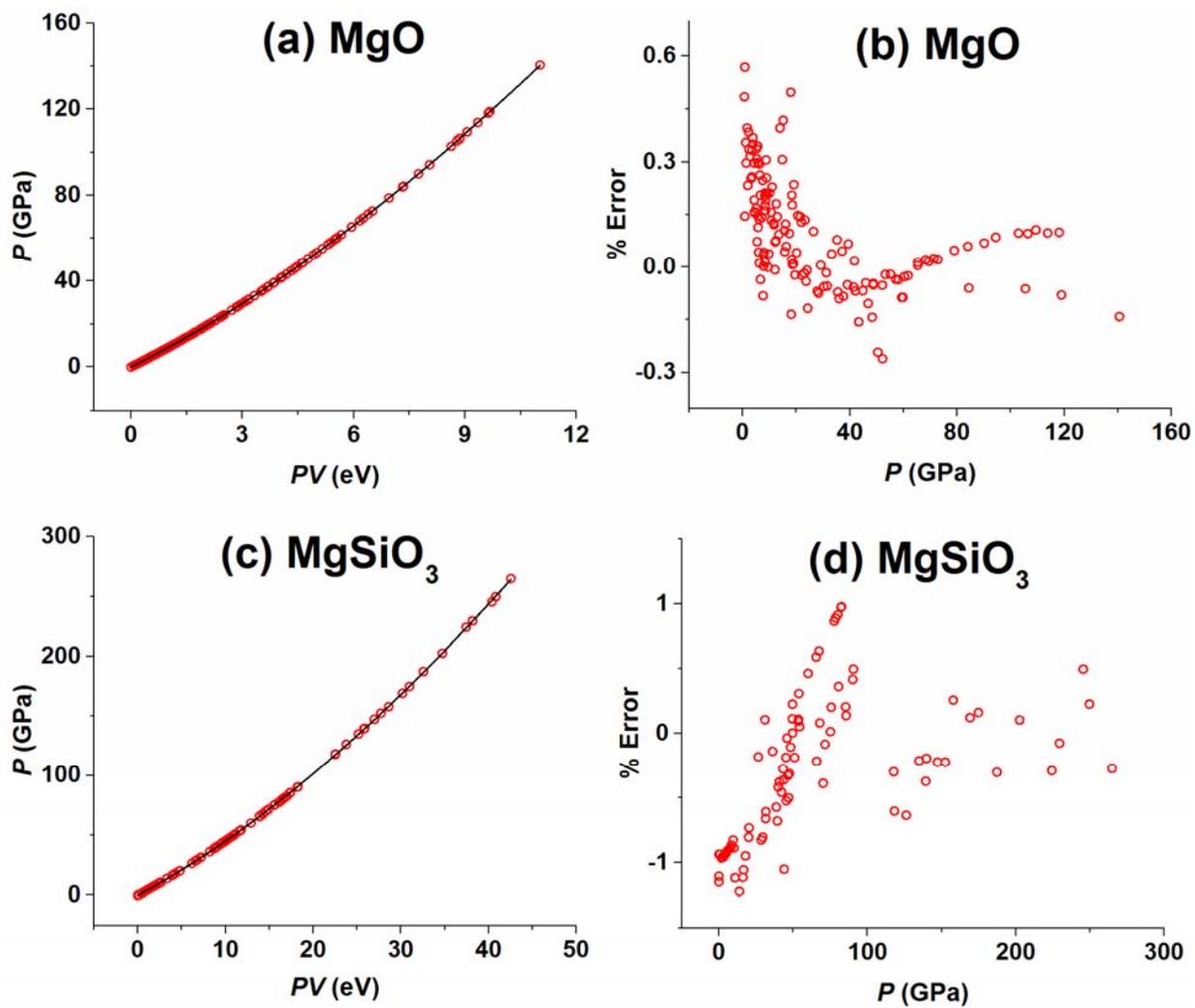

Figure 2

Supplementary Information

for

**Condensed-matter equation of states covering a wide region of pressure studied experimentally**

Elijah E. Gordon[1], Jürgen Köhler[2] & Myung-Hwan Whangbo[1,*]

[1] Department of Chemistry, North Carolina State University, Raleigh, NC 27695-8204, USA

[2] Max-Planck-Institut für Festkörperforschung, D-70569 Stuttgart, Germany

* Corresponding author, email: mike_whangbo@ncsu.edu

## 1. Results of the DFT calculations leading to the EOS, Eq. 1

### (a) DFT calculations for the structures of Te under various pressures

The space groups of the known atomic structures of elemental Te at various pressures $P$ and temperature $T$, their volumes $V$ per atom, the calculated energies $E$ per atom as well as the energy $PV$

| $P$ (GPa) | $T$ | Space Group (#) | $E$/Atom (eV) | $V$/Atom (Å$^3$) | $PV$/Atom (eV) |
|---|---|---|---|---|---|
| 0.0001 | 298 K | P3(1)21 (152)[1] | -3.13 | 33.94 | 0.00 |
| 4.5 | 298 K | P2(1) (4)[2] | -2.91 | 27.76 | 0.78 |
| 8 | 298 K | Cm (12)[3] | -2.91 | 26.49 | 1.32 |
| 23 | 473 K | R-3m (166)[4] | -2.51 | 22.50 | 3.23 |
| 33 | 298 K | Im-3m (229)[5] | -2.15 | 20.65 | 4.25 |

References

1. Adenis, C.; Langer, V.; Lindqvist, O. *Acta Crystallogr. C* **1989**, 45, 941-942.

2. Aoki, K.; Shimomura, O.; Minomura, S. *J. Phys. Soc. Jpn.* **1980**, 48, 551-556.

3. Takumi, M.; Masamitsu, T.; Nagata, K. *J. Phys.- Condens. Mat.* **2002**, 14, 10609-10613.

4. Hejny, C.; Falconi, S.; Lundegaard, L. F.; McMahon, M. I. *Phys. Rev. B* **2006**, 74, 174119.

5. Parthasarathy, G.; Holzapfel, W. B. *Phys. Rev. B* **1988**, 37, 8499-8501.

**(b) DFT calculations for the structures of Se under various pressures**

The space groups of the known atomic structures of elemental Se at various pressures $P$ and temperature $T$, their volumes $V$ per atom, the calculated energies $E$ per atom as well as the energy $PV$

| $P$ (GPa) | $T$ | Space Group (#) | $E$/Atom (eV) | $V$/Atom (Å$^3$) | $PV$/Atom (eV) |
|---|---|---|---|---|---|
| 0.0001 | 298 K | P2(1)/c (14)[1] | -3.47 | 30.13 | 0.00 |
| 4.6 | 298 K | P3(1)21 (152)[2] | -3.30 | 22.67 | 0.65 |
| 23 | 298 K | P2(1) (4)[3] | -2.88 | 17.37 | 2.49 |
| 87.9 | 298 K | R-3m (166)[3] | -1.53 | 12.93 | 7.09 |
| 140 | 298 K | Im-3m (229)[3] | -0.22 | 11.25 | 9.83 |

References

1. Marsh, R. E.; Pauling, L.; McCullough, J. D. *Acta Crystallogr.* **1953**, 6, 71.

2. Parthasarathy, G.; Holzapfel, W. B. *Phys. Rev. B* **1988**, 38, 10105.

3. Akahama, Y.; Kobayashi, M.; Kawamura, H. *Phys. Rev. B* **1993**, 47, 20-26.

**(c) DFT calculations for the structures of S under various pressures**

The space groups of the known atomic structures of elemental S at various pressures $P$ and temperature $T$, their volumes $V$ per atom, the calculated energies $E$ per atom as well as the energy $PV$

| $P$ (GPa) | $T$ | Space Group (#) | $E$/Atom (eV) | $V$/Atom (Å³) | $PV$/Atom (eV) |
|---|---|---|---|---|---|
| 0.0001 | 298 K | Fddd (70)[1] | -4.07 | 25.97 | 0.00 |
| 3 | 673 K | P3(2)21 (154)[2] | -4.00 | 20.81 | 0.39 |
| 12 | 298 K | I4(1)/acd (142)[3] | -3.78 | 16.7 | 1.25 |
| 160 | 298 K | R-3m (166)[4] | -1.07 | 8.77[a] (8.87) | 8.85 |
| 173 | 298 K | R-3m (166)[5] | -0.76 | 8.50[a] (10.68) | 9.18 |
| 206.5 | 298 K | R-3m (166)[6] | -0.20 | 8.01 | 10.32 |

[a] We employed the volumes optimized by DFT calculations, because the experimental value at 173 GPa should not be greater than that at 160 GPa (the numbers in the parentheses).

References

1. Warren, B. E.; Burwell, J. T. *J. Chem. Phys.* **1935**, 3, 6-8.

2. Crichton, W. A.; Vaughan, G. B. M.; Mezouar, M., *Z. Kristallogr.* **2001**, 216, 417-419.

3. Degtyareva, O.; Gregoryanz, E.; Somayazulu, M.; Dera, P.; Mao, H.-K.; Hemley, R. J., *Nat. Mater.* **2005**, 4, 152-155.

4. Degtyareva, O.; Gregoryanz, E.; Somayazulu, M.; Dera, P.; Mao, H.-K.; Hemley, R. J., *Phys. Rev. B* **2005**, 71, 214104.

**(d) DOS plots calculated for Te under various pressures**

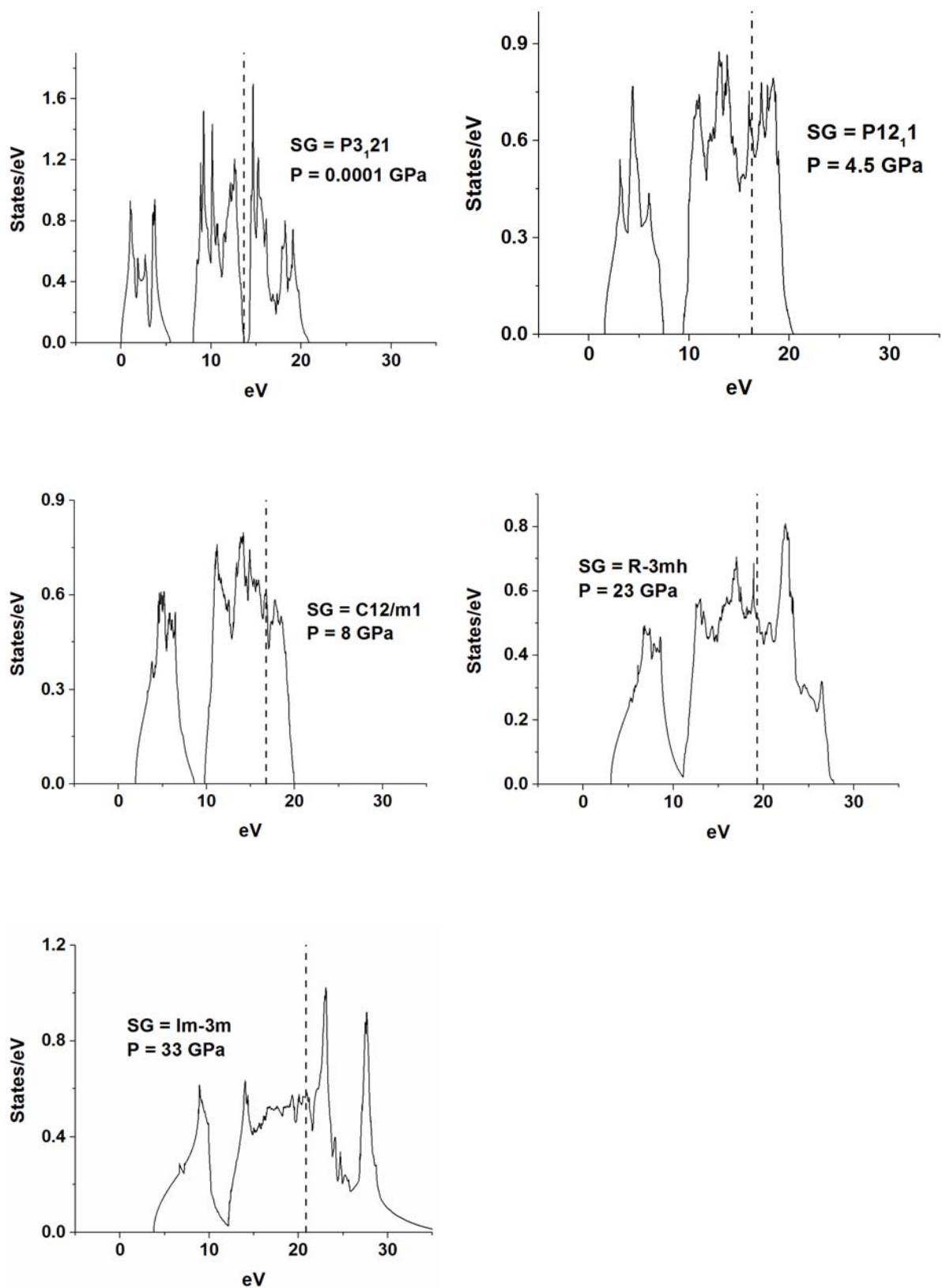

**(e) DOS plots calculated for Se under various pressures**

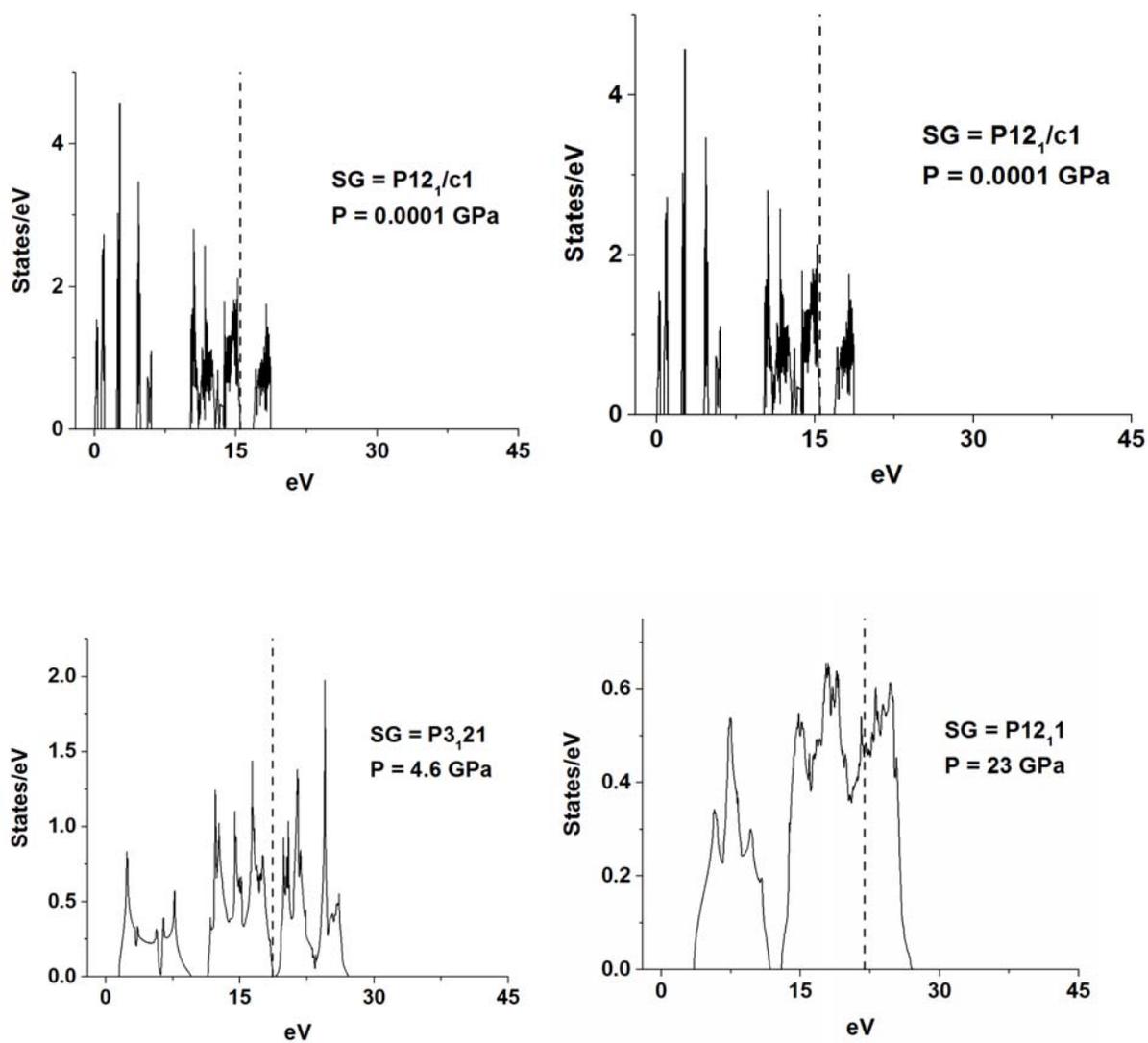

**(f) DOS plots calculated for S under various pressures**

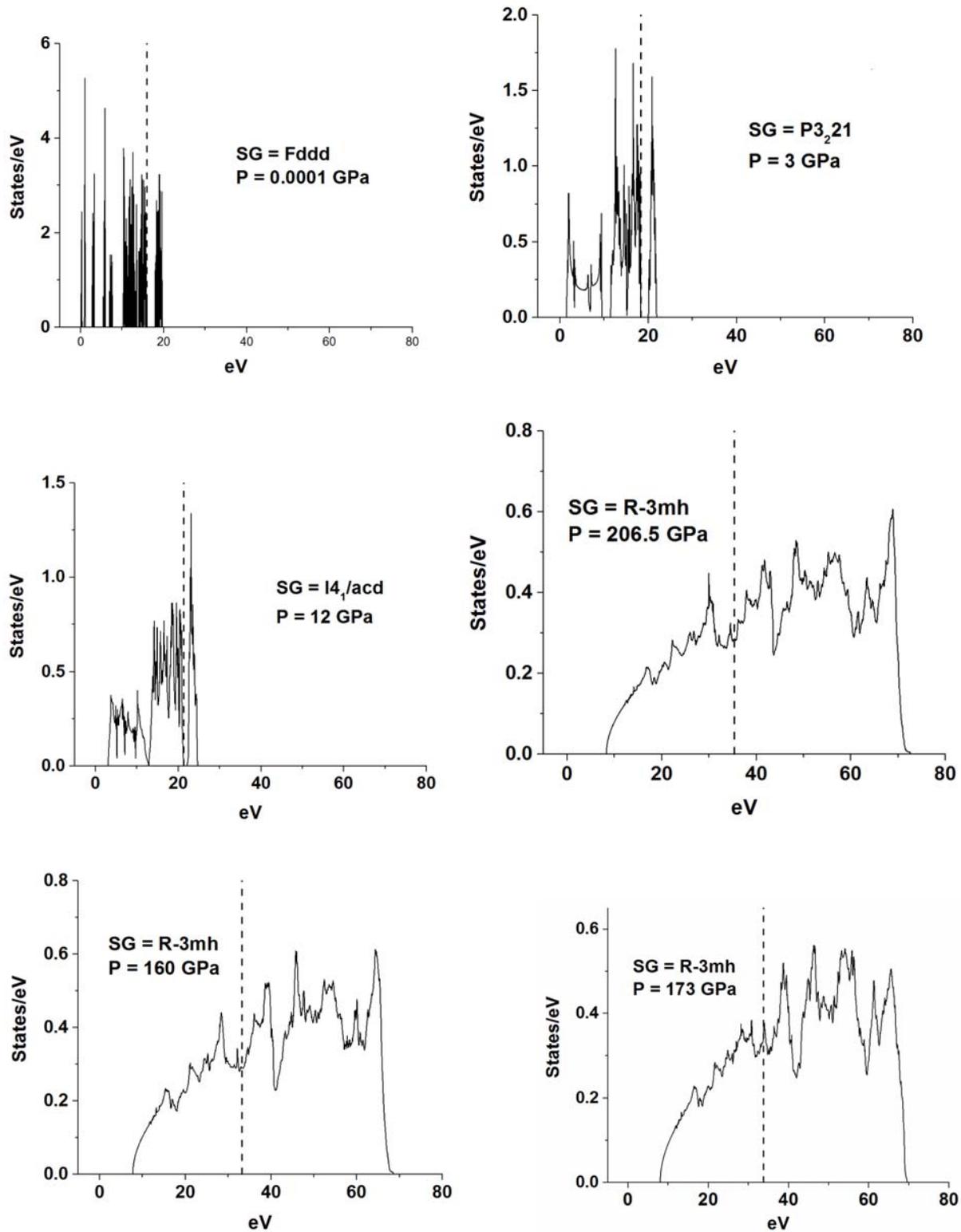

## 2. Analyses of the empirical EOSs for chalcogen

(a) Energy-vs-$P$, $P$-vs-$PV$, % error-vs-$P$, and $B(P)$-vs-$P$ plots obtained for Se

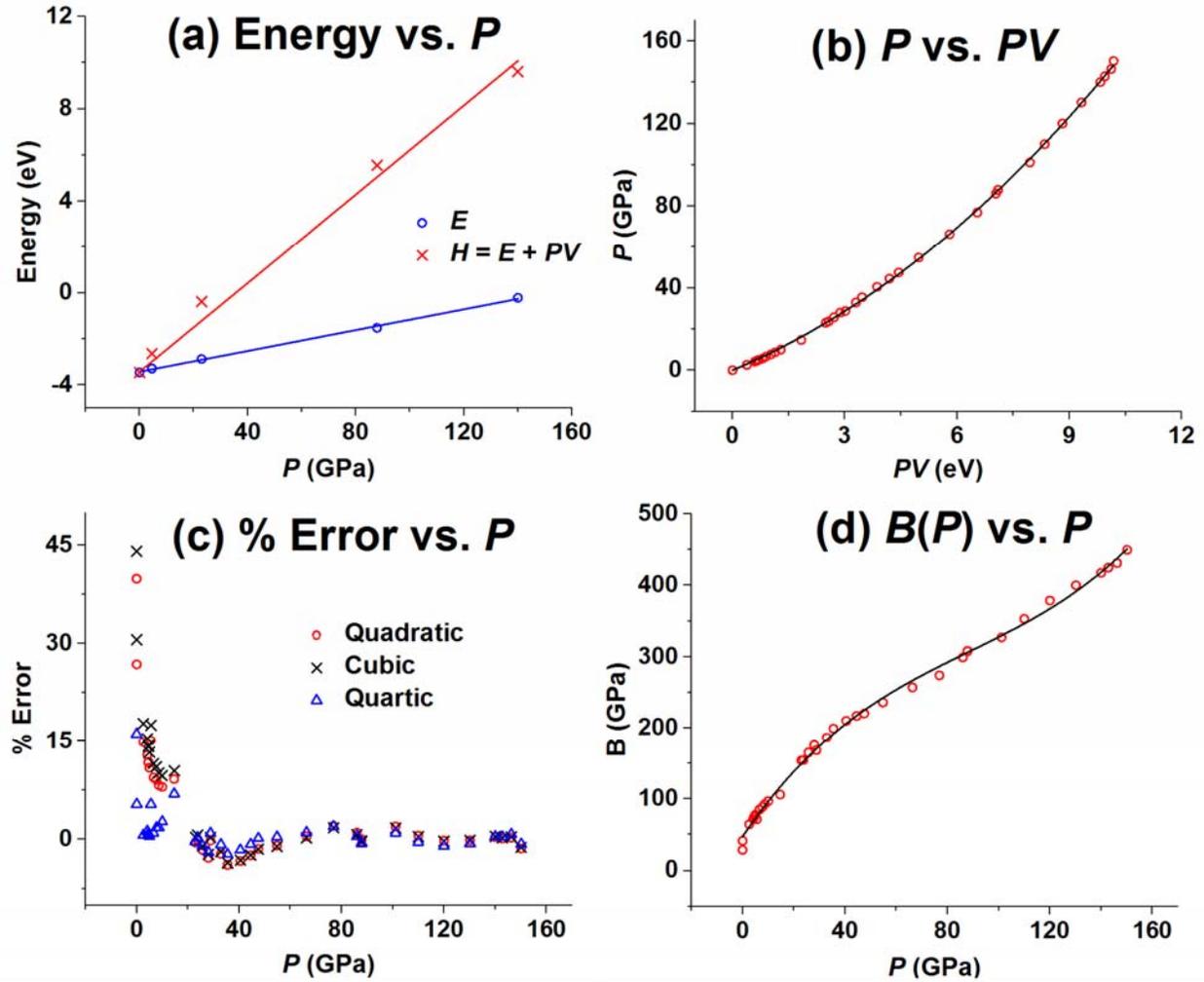

a) The $E$ vs. $P$ and $H$ vs. $P$ plots calculated for Se, where $E$, $H$ and $PV$ are in units of eV. The fitting coefficients $a_1$, $a_0$ and $b_1$ for $E = a_1 P + a_0$ and $H = b_1 P + a_0$ are -3.2146, 0.0206 and 0.0945, respectively. b) The $P$ vs. $PV$ plot obtained for Se, where the solid line is the fitting curve obtained from the cubic approximation. c) The pressure-dependence of the % error, $100 \times (P_{calc} - P_{expt})/P_{expt}$, obtained for Se from the cubic approximation. d) The pressure-dependence of the bulk modulus $B(P)$ calculated for Se.

**(b) Energy-vs-*P*, *P*-vs-*PV*, % error-vs-*P*, and *B*(*P*)-vs-*P* plots obtained for S**

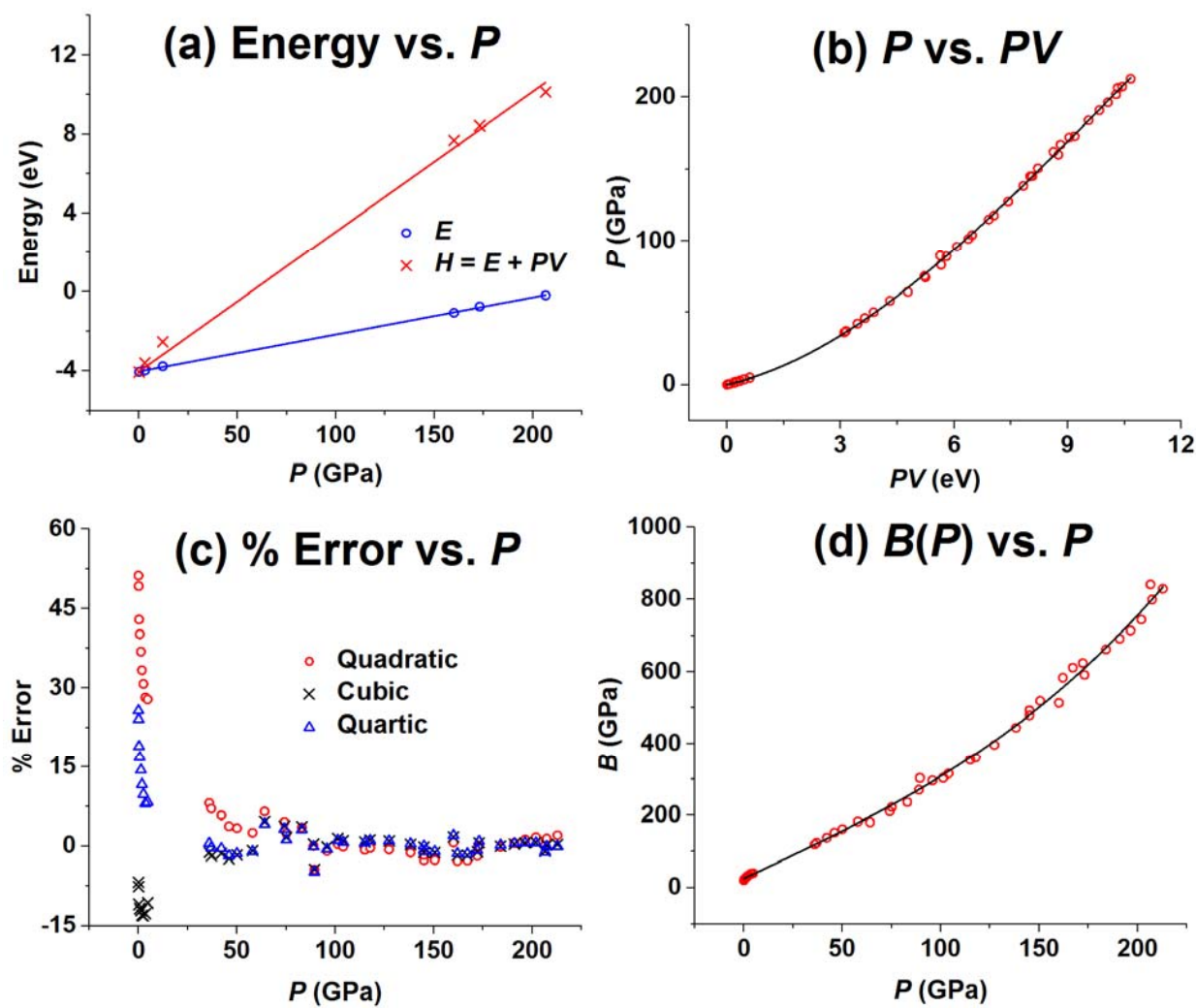

a) The $E$ vs. $P$ and $H$ vs. $P$ plots calculated for S, where $E$, $H$ and $PV$ are in units of eV. The fitting coefficients $a_1$, $a_0$ and $b_1$ for $E = a_1 P + a_0$ and $H = b_1 P + a_0$ are -3.8808, 0.0175 and 0.0696, respectively. b) The $P$ vs. $PV$ plot obtained for S, where the solid line is the fitting curve obtained from the cubic approximation. c) The pressure-dependence of the % error, $100 \times (P_{calc} - P_{expt})/P_{expt}$, obtained for S from the cubic approximation. d) The pressure-dependence of the bulk modulus $B(P)$ calculated for S.

**(c) The fitting of the *P*-vs-*V* data for chalcogens, used for DFT calculations, with the equation $P = \alpha_1(PV) + \alpha_2(PV)^2$**

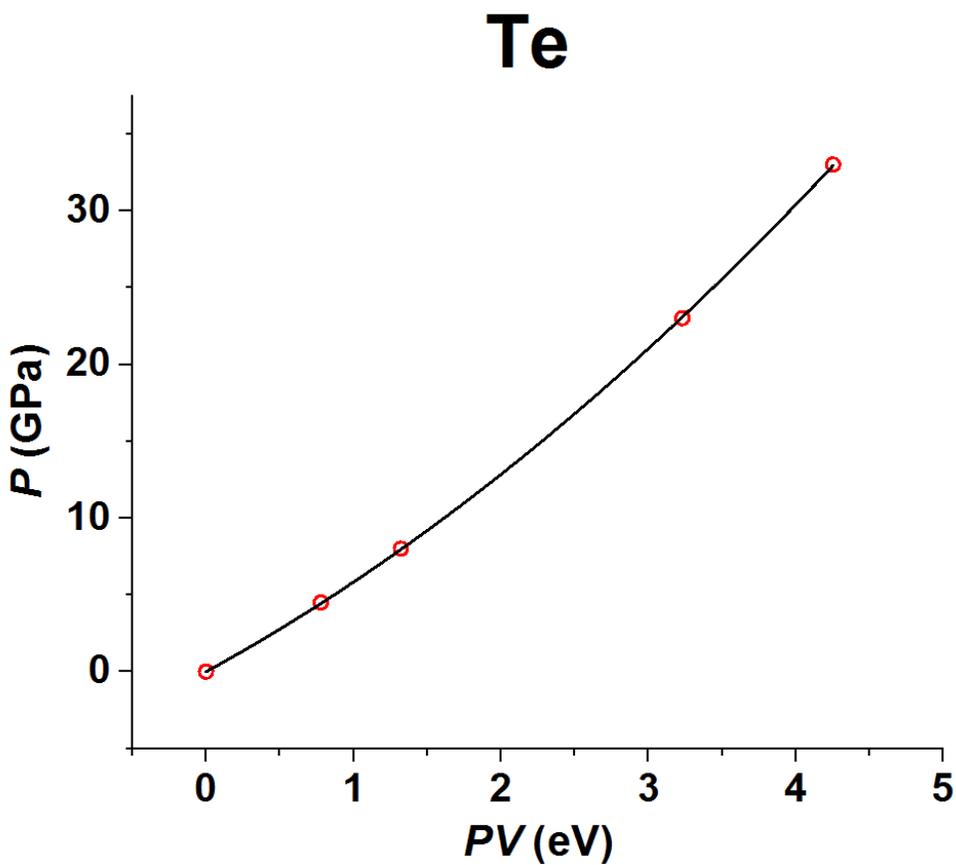

For Te: $\alpha_1 = 5.245$ and $\alpha_2 = 0.5886$

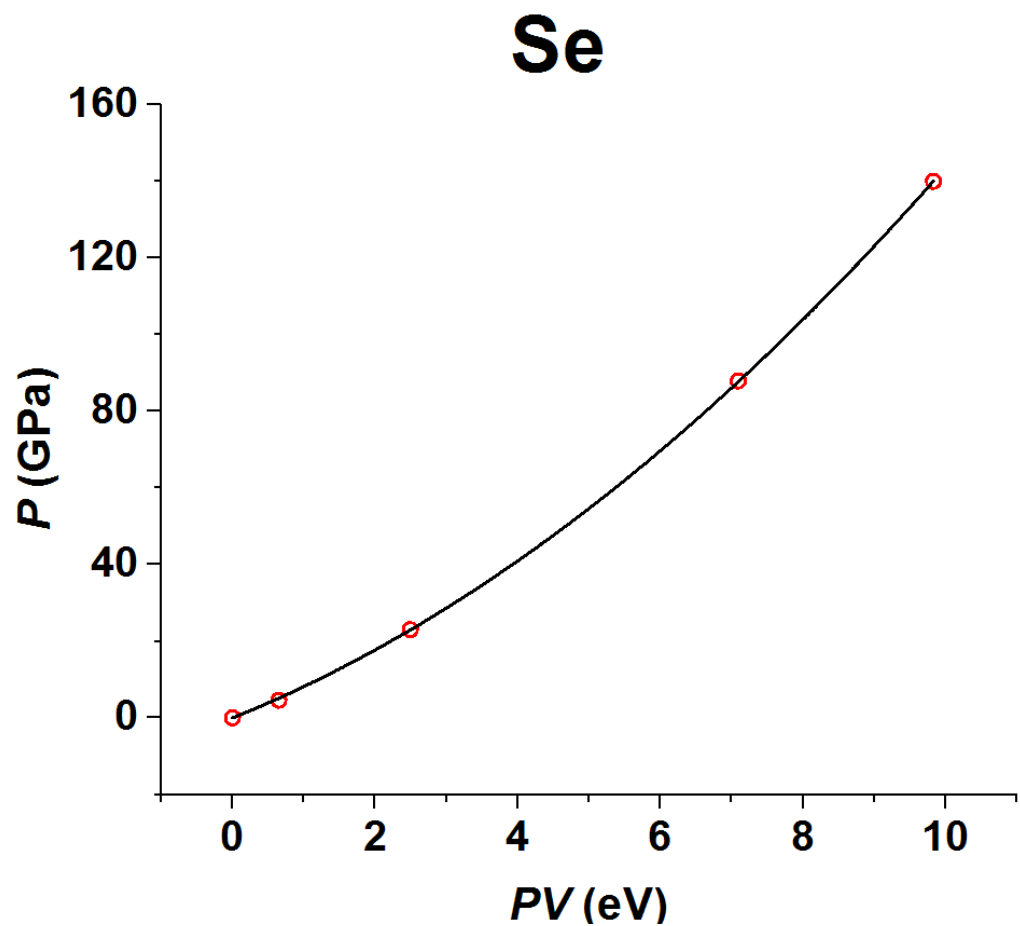

For Se: $\alpha_1 = 7.500$ and $\alpha_2 = 0.6866$

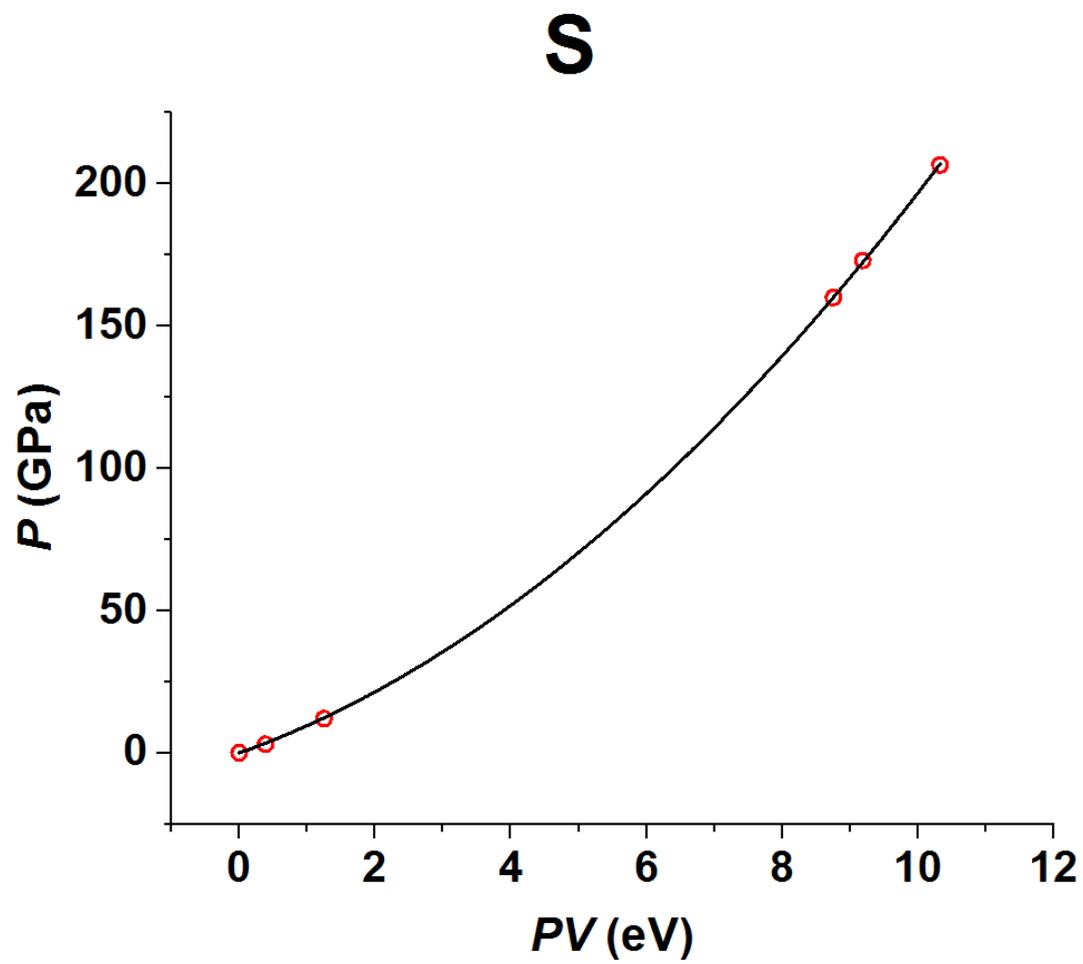

For S: $\alpha_1 = 8.471$ and $\alpha_2 = 1.121$

**(d) The coefficients $\alpha_i$ of the isothermal EOS, $P = \alpha_1(PV) + \alpha_2(PV)^2 + \alpha_3(PV)^3 + \ldots$, for elemental chalcogens in the quadratic, cubic and quartic approximations, obtained by using $P$ in GPa units and $V$ in Å$^3$ units.**

1) For Te from the 0 – 330 GPa region

| Approximation | $\alpha_1$ | $\alpha_2$ | $\alpha_3$ | $\alpha_4$ |
|---|---|---|---|---|
| Quadratic | 0.04491 | 0.9036×10$^{-5}$ | | |
| Cubic | 0.03785 | 1.501×10$^{-5}$ | -1.130×10$^{-9}$ | |
| Quartic | 0.03769 | 1.527×10$^{-5}$ | -1.244×10$^{-9}$ | 1.523×10$^{-14}$ |

2) For Se from the 0 – 150 GPa region

| Approximation | $\alpha_1$ | $\alpha_2$ | $\alpha_3$ | $\alpha_4$ |
|---|---|---|---|---|
| Quadratic | 0.04644 | 2.712×10$^{-5}$ | | |
| Cubic | 0.04782 | 2.421×10$^{-5}$ | 1.325×10$^{-9}$ | |
| Quartic | 0.03856 | 5.998×10$^{-5}$ | -37.34×10$^{-9}$ | 1.251×10$^{-11}$ |

3) For S from the 0 – 213 GPa region

| Approximation | $\alpha_1$ | $\alpha_2$ | $\alpha_3$ | $\alpha_4$ |
|---|---|---|---|---|
| Quadratic | 0.05824 | 4.035×10$^{-5}$ | | |
| Cubic | 0.03587 | 7.987×10$^{-5}$ | -1.620×10$^{-8}$ | |
| Quartic | 0.04840 | 4.250×10$^{-5}$ | +1.820×10$^{-8}$ | -9.942×10$^{-12}$ |

**(e) Pressure-dependence of the absolute errors, $\Delta P = P_{calc} - P_{expt}$, calculated for Te, Se and S**

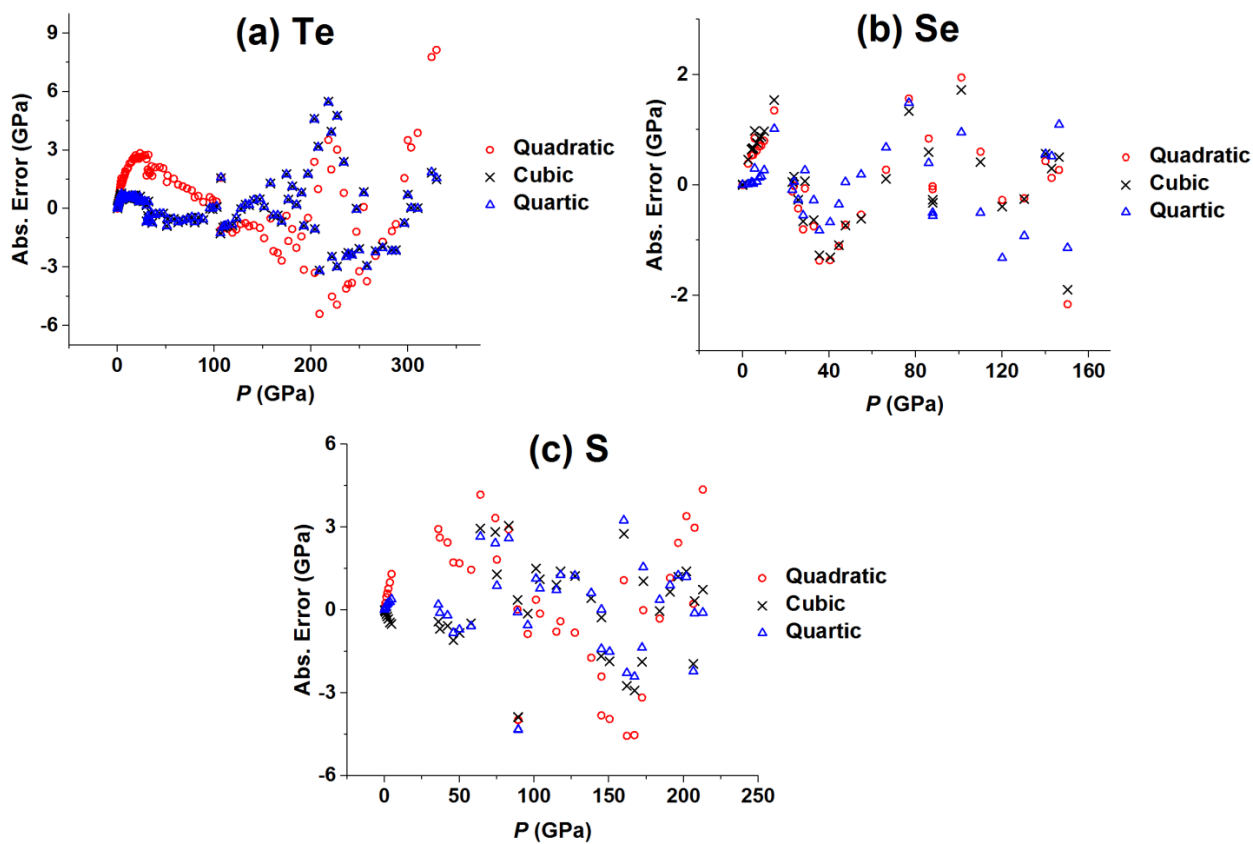

The pressure-dependence of the absolute error, $(P_{calc} - P_{expt})$, obtained for (a) Te, (b) Se and (c) S from the cubic approximation.

**3. Analyses of the empirical EOSs obtained for various condensed matter listed in Table 1**

In this section we show the *P*-vs. *PV*, the % error-vs-*P*, and the *B*(*P*)-vs-*P* plots for the various condensed matter studied in this work. The references (a) – (q) were defined in Tables 1 and 3.

**(a) Sn**

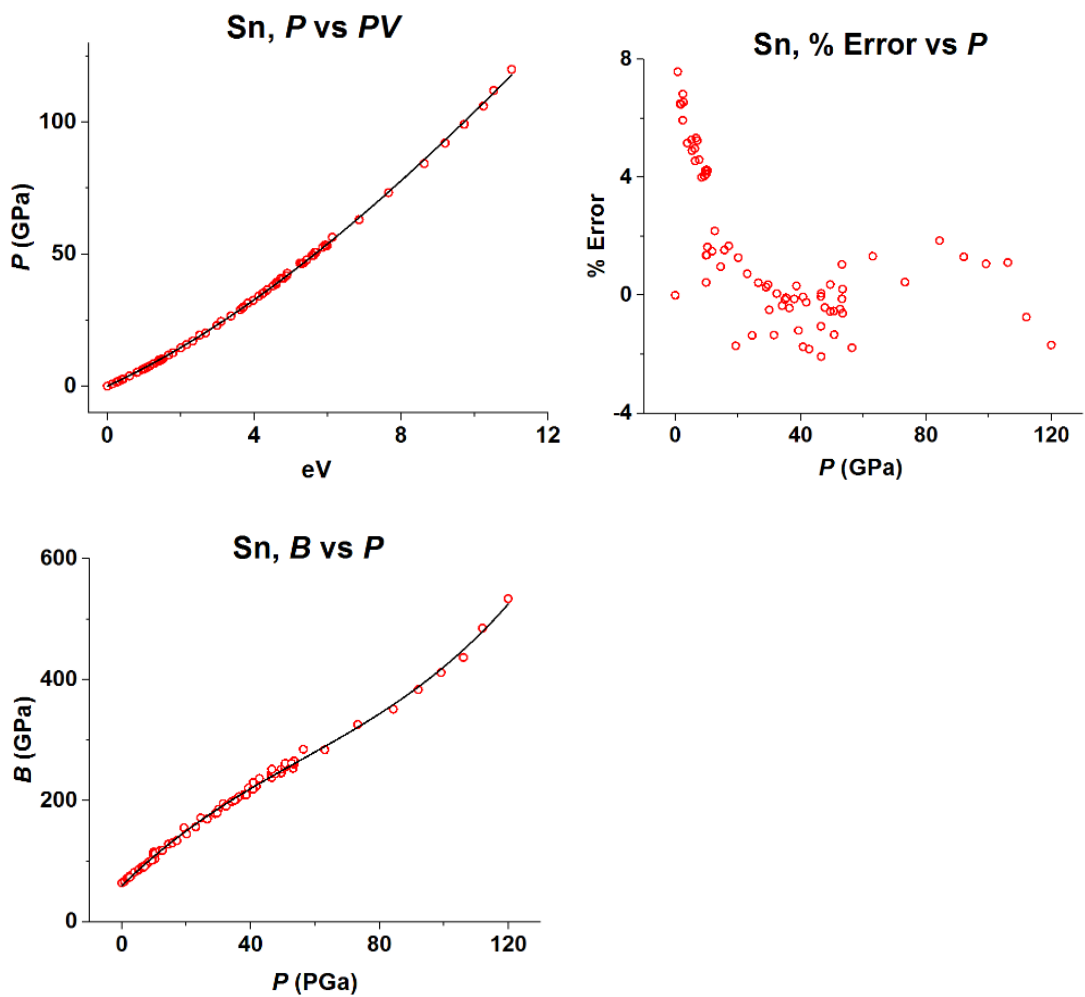

References

(a) Liu, M., & Liu, L., Compressions and phase transitions of tin to half a megabar. *High Temp.-High Press*. **18**, 79-85 (1986)

(b) Olijnyk, H., & Holzapfel, W. B., Phase transitions in Si, Ge and Sn under pressure. *J. Phys.-Paris Colloq*. **45**, Suppl. 11, C8-153 (1984).

(c) Desgreniers, S., Vohra, Y. K., & Ruoff, A. L., Tin at high pressure: An energy-dispersive x-ray diffraction study to 120 GPa. *Phys. Rev. B* **39**, 10359-10361 (1988).

(o) Kamioka, H., Temperature Variations of Elastic Moduli up to Eutectic Temperature in Tin-Bismuth Alloys, *Jpn. J. Appl. Phys*. **22**, 1805 (1983).

(p) Vaiyda, S. N., Kennedy, G. C., Compressibility of 22 elemental solids to 45 KB, *J. Phys. Chem. Solids* **31**, 2329 (1970).

**(b) Au**

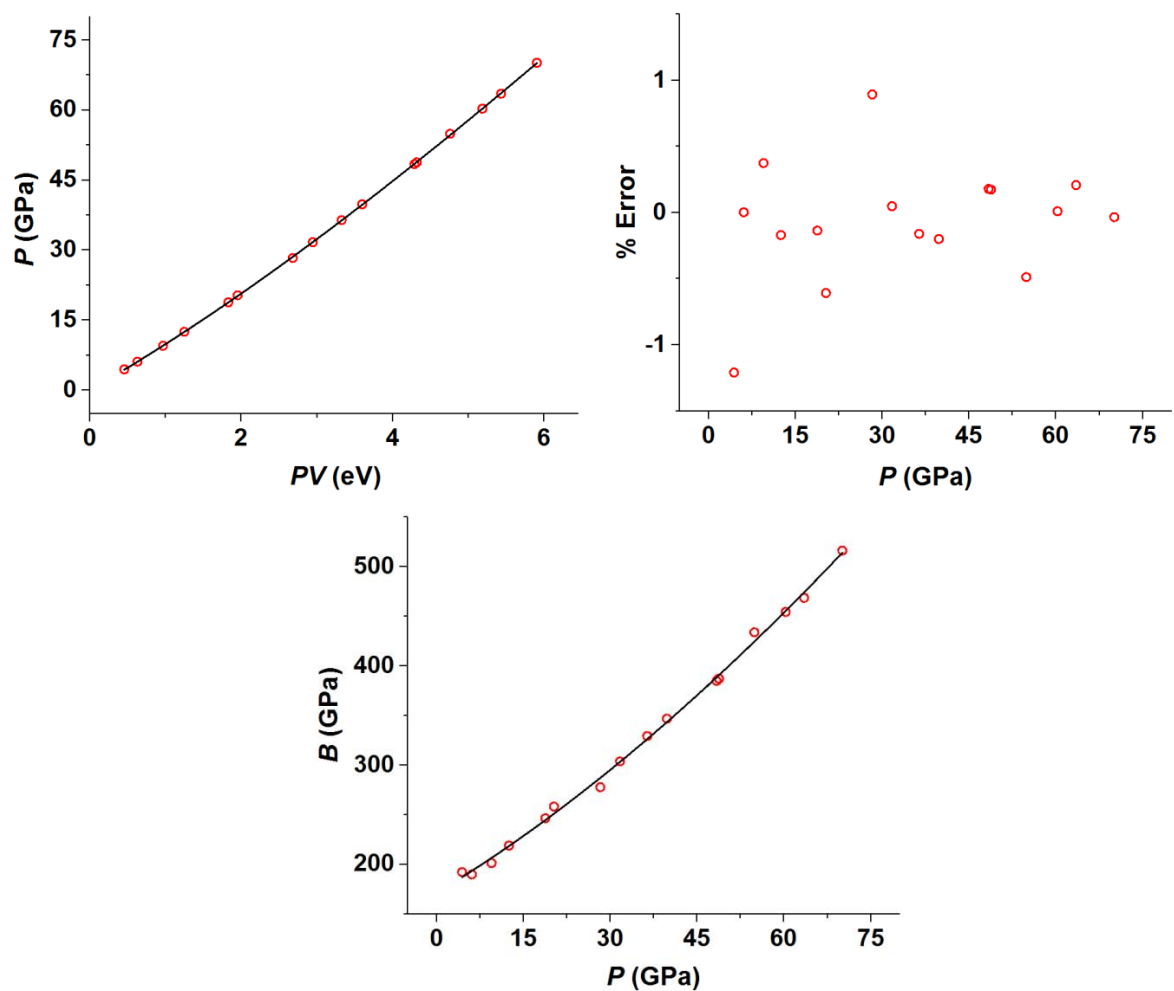

References

(d) Heinz, D. L. & Jeanloz, R., The equation of state of the gold calibration standard. *J. Appl. Phys.* **55**, 885-893 (1984).

**(c) Cu**

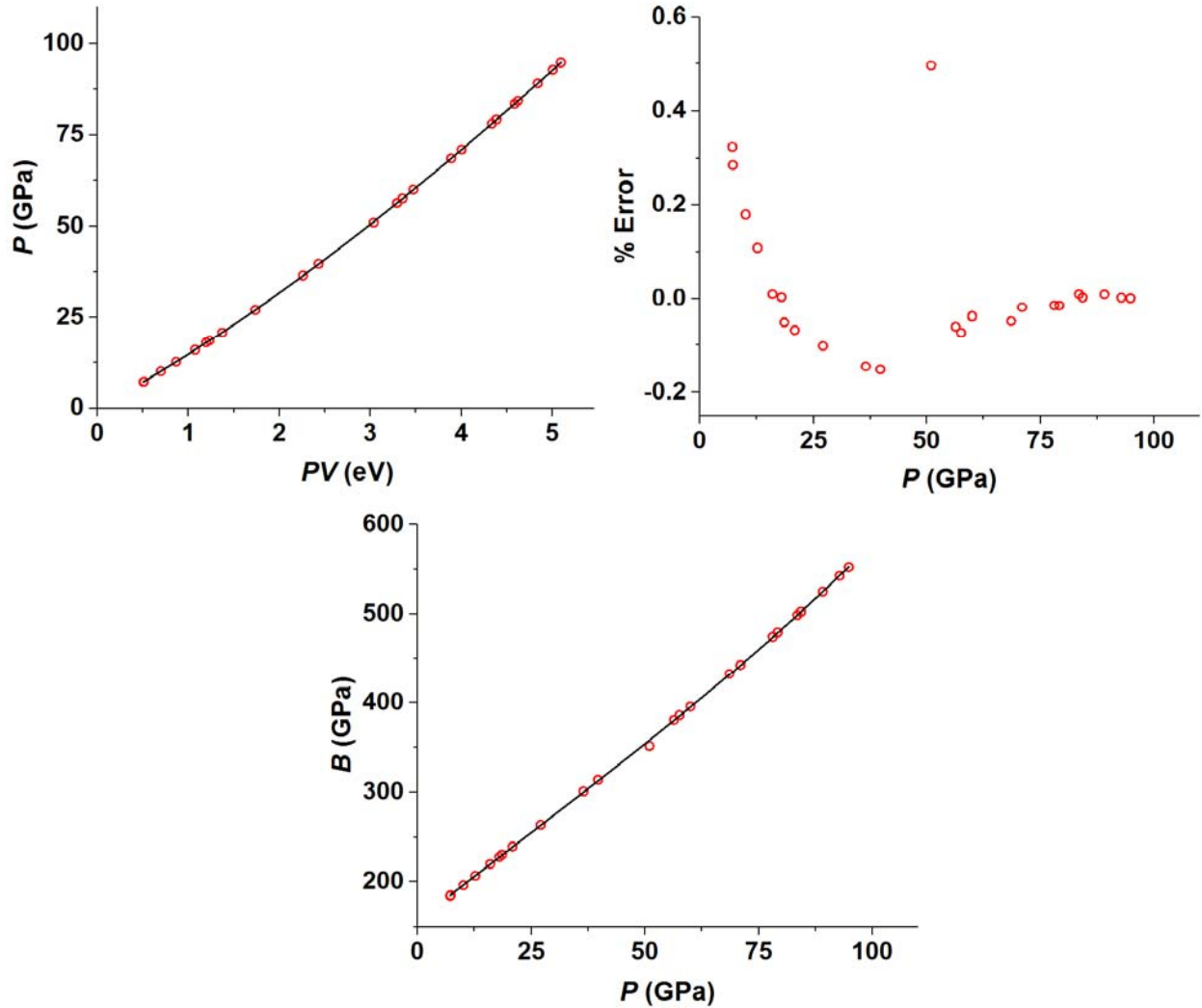

**(d) LiF**

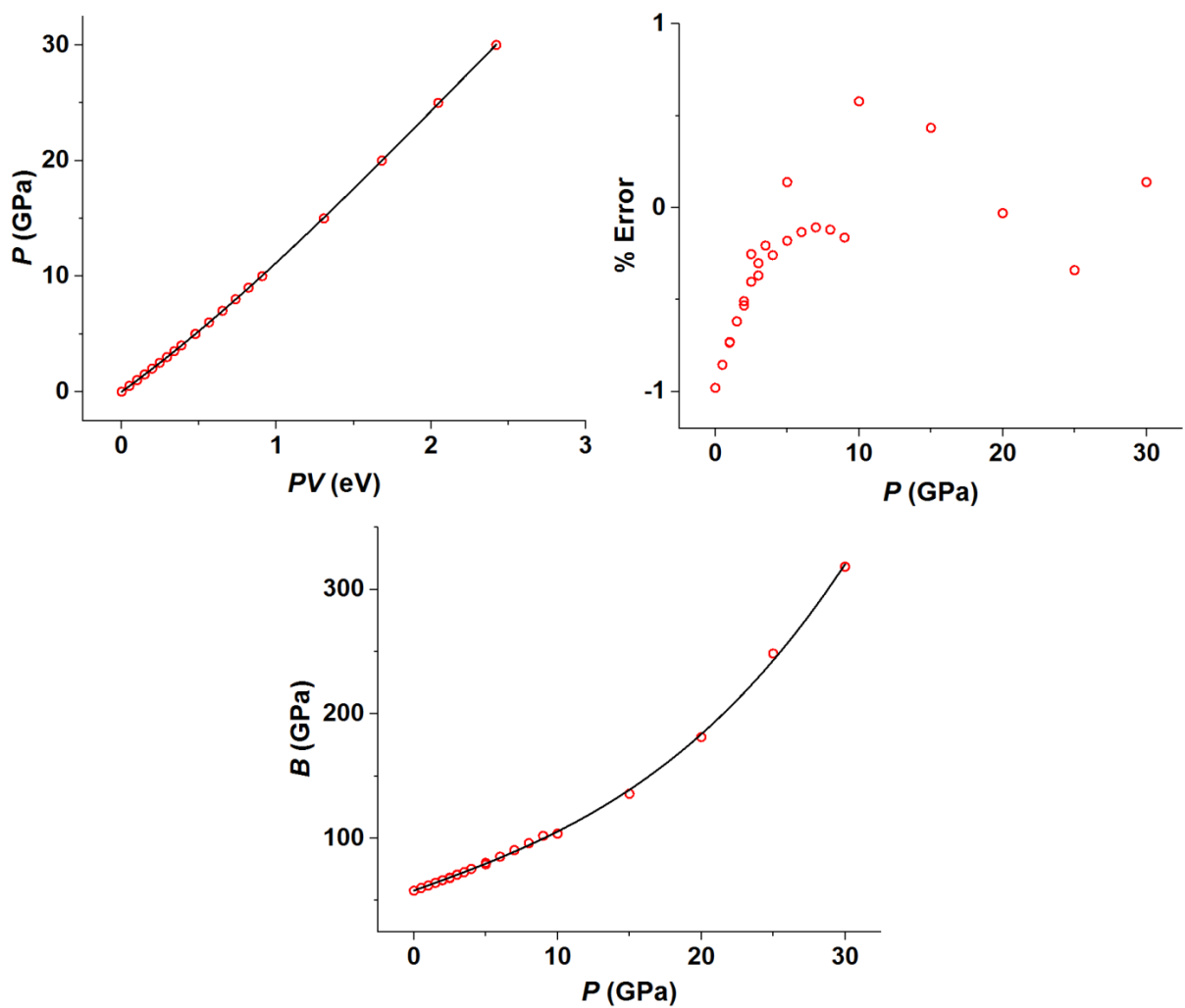

**(e) NaF**

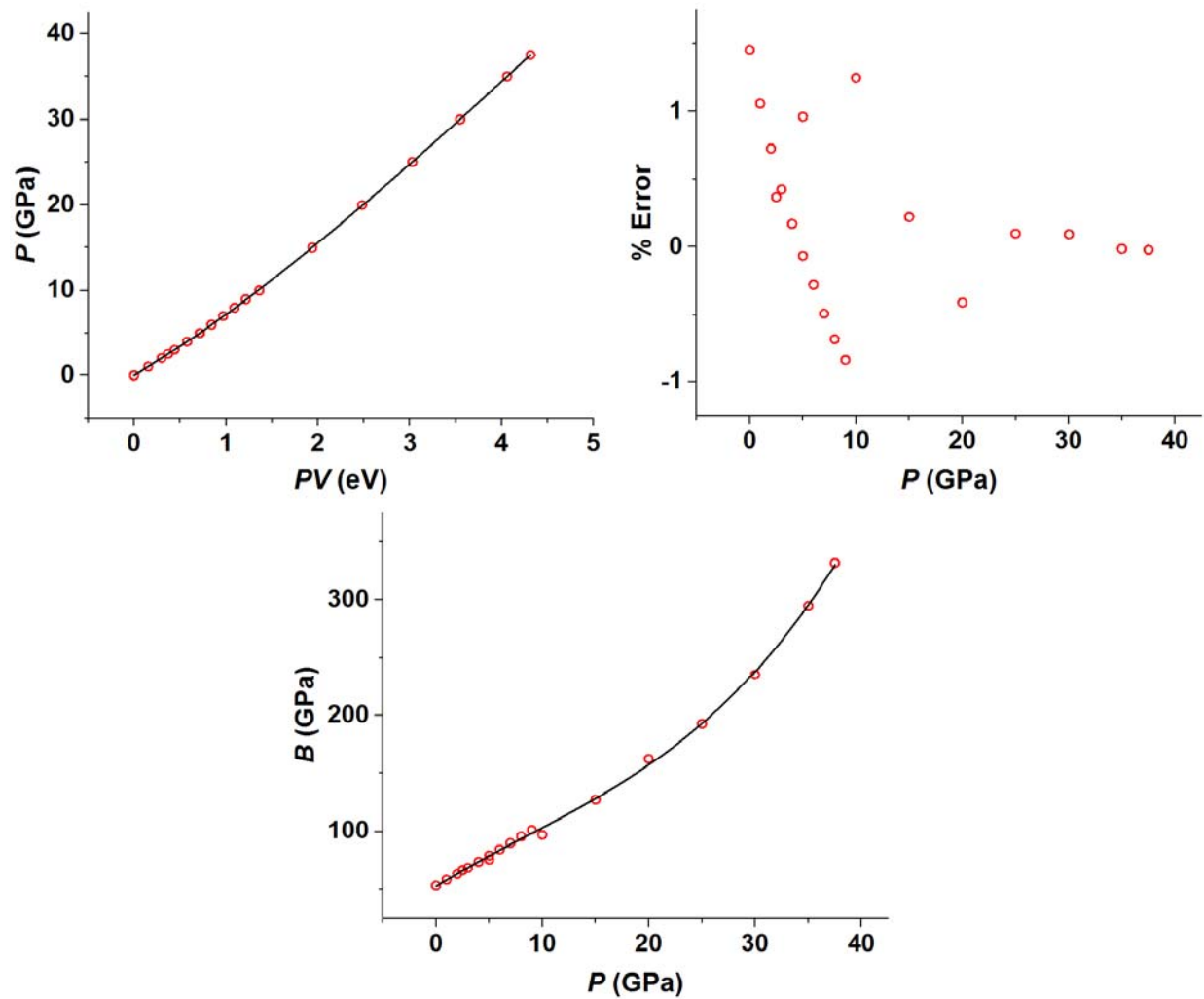

References

(f) Yagi, T. Experimental determination of thermal expansivity of several alkali halides at high pressure. *J. Phys. Chem. Solids* **39**, 563-571 (1978).

(g) Pagannone, M. & Drickamer, H. G., Effect of high pressure on the compressibilities of NaI, LiF, and NaF. *J. Chem. Phys.* **43**, 2266-2268 (1965).

(q) Barsch, G. R. & Chang, Z. P., Adiabatic, Isothermal, and Intermediate Pressure Derivatives of the Elastic Constants for Cubic Symmetry. *Phys. Status Solidi B* **19**, 139-151 (1967).

**(f) NaCl**

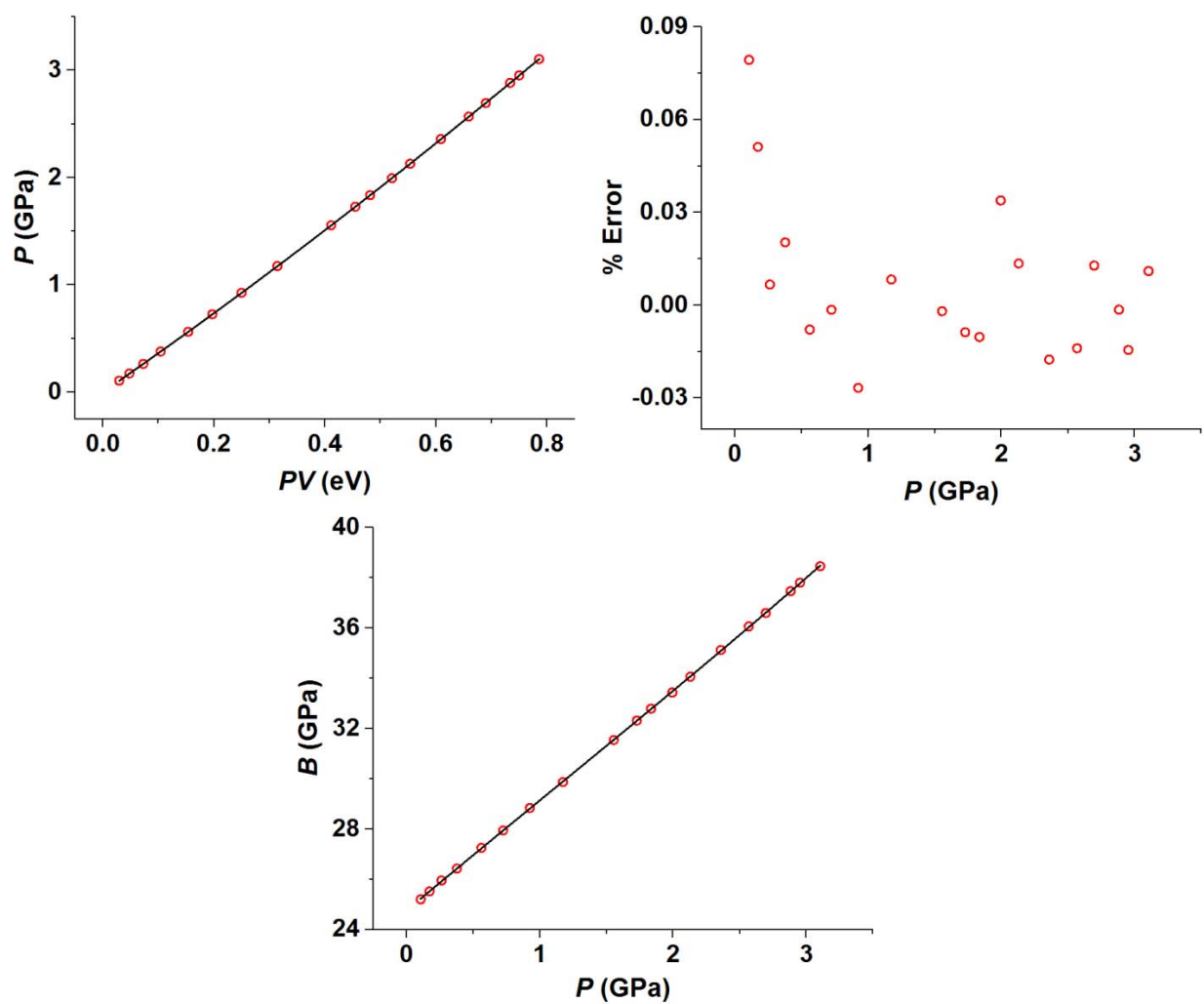

References

(i) Boehler, R., & Kennedy, G. C., Equation of state of sodium chloride. *J. Phys. Chem. Solids* **41**, 517-523 (1980).

(q) Barsch, G. R. & Chang, Z. P., Adiabatic, Isothermal, and Intermediate Pressure Derivatives of the Elastic Constants for Cubic Symmetry. *Phys. Status Solidi B* **19**, 139-151 (1967).

**(g) CsCl**

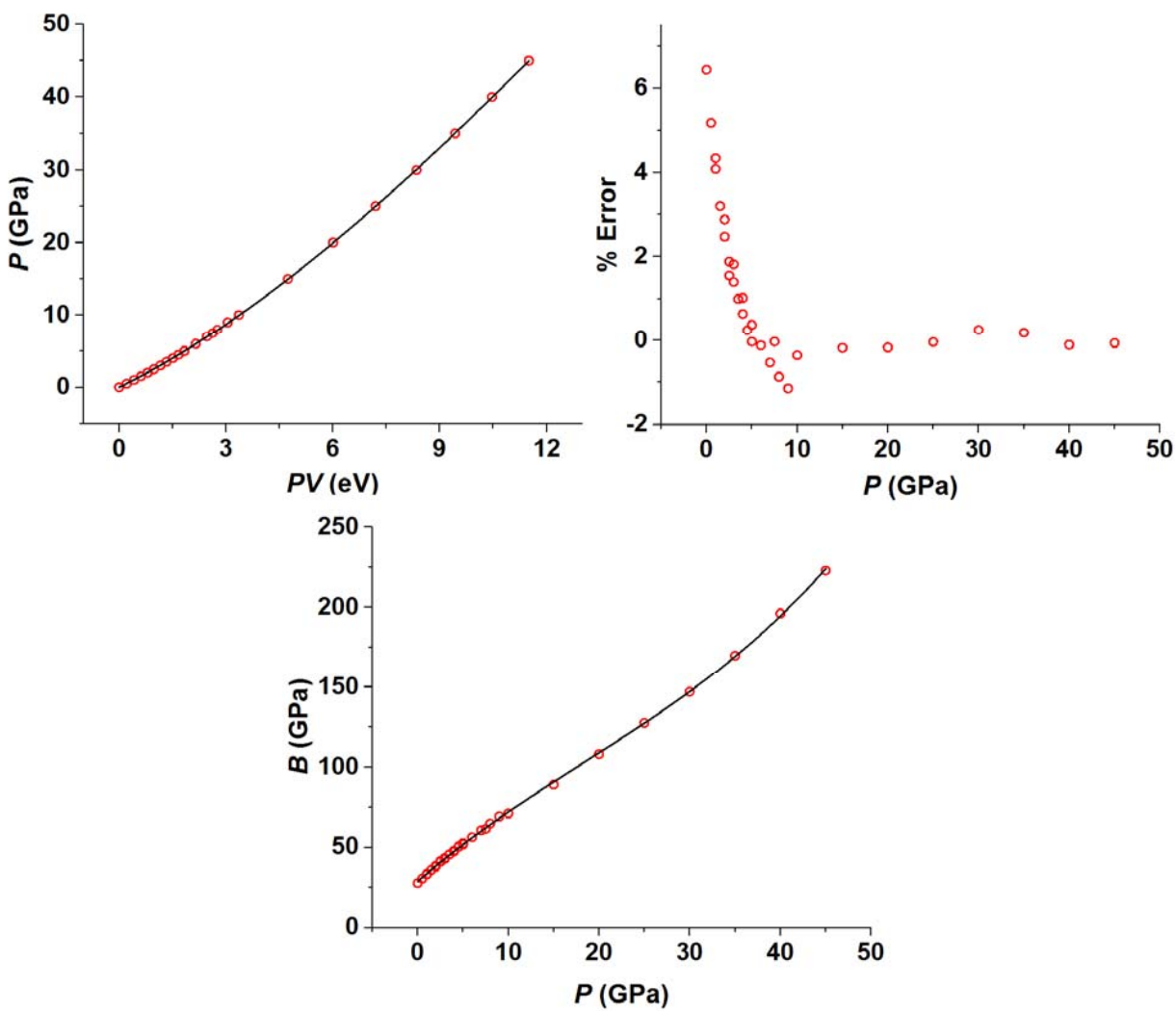

References

(f) Yagi, T. Experimental determination of thermal expansivity of several alkali halides at high pressure. *J. Phys. Chem. Solids* **39**, 563-571 (1978).

(j) Perez-Albuerne, E. A., & Drickamer, H. G., Effect of high pressures on the compressibilities of seven crystals having the NaCl or CsCl structure. *J. Chem. Phys.* **43**, 1381-1387 (1965).

(k) Vaidya, S. N. & Kennedy, G. C., Compressibility of 27 halides to 45 kbar. *J. Phys. Chem. Solids* **32**, 951-964 (1971).

**(h) Ice VII**

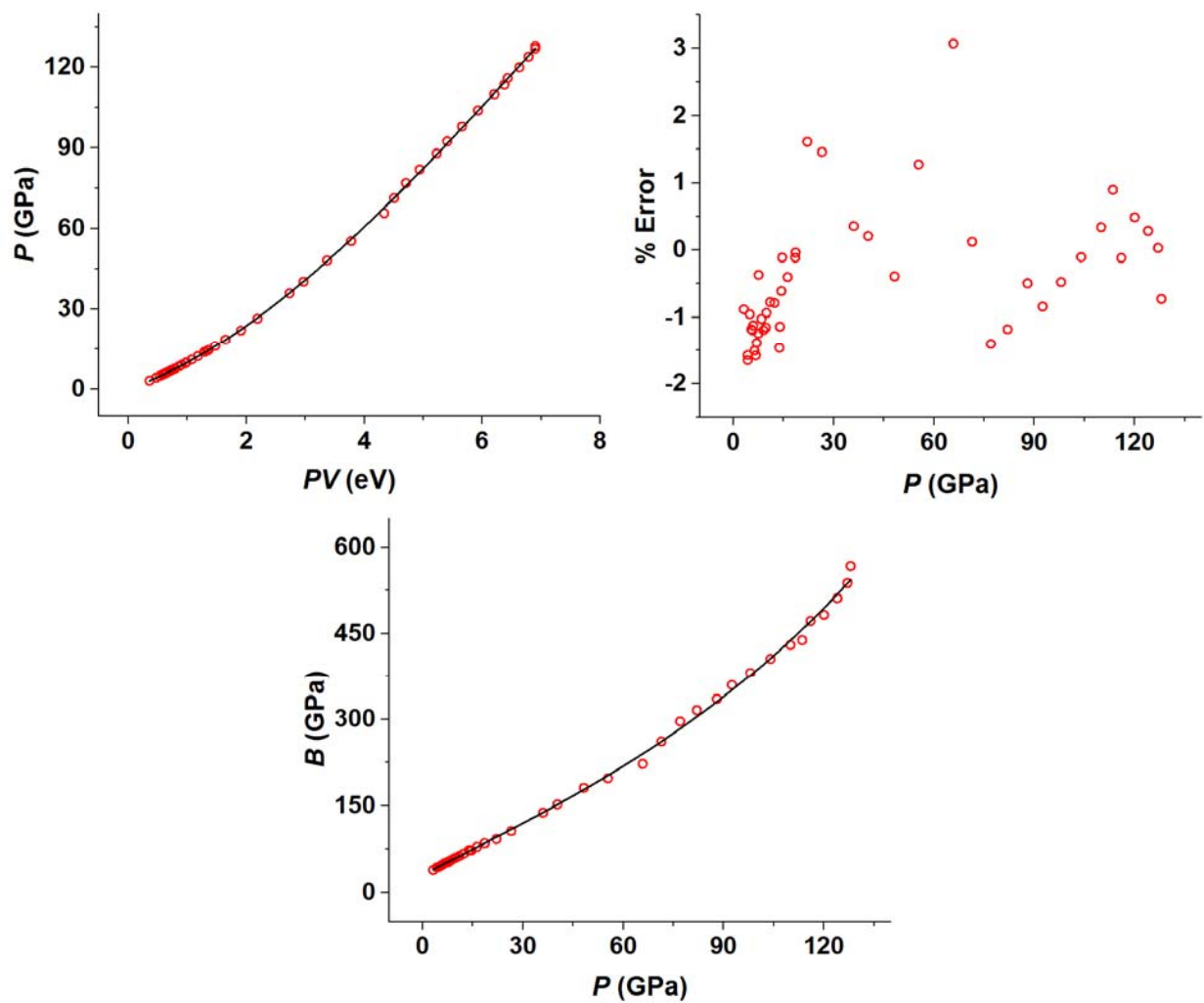

**(i) Ar**

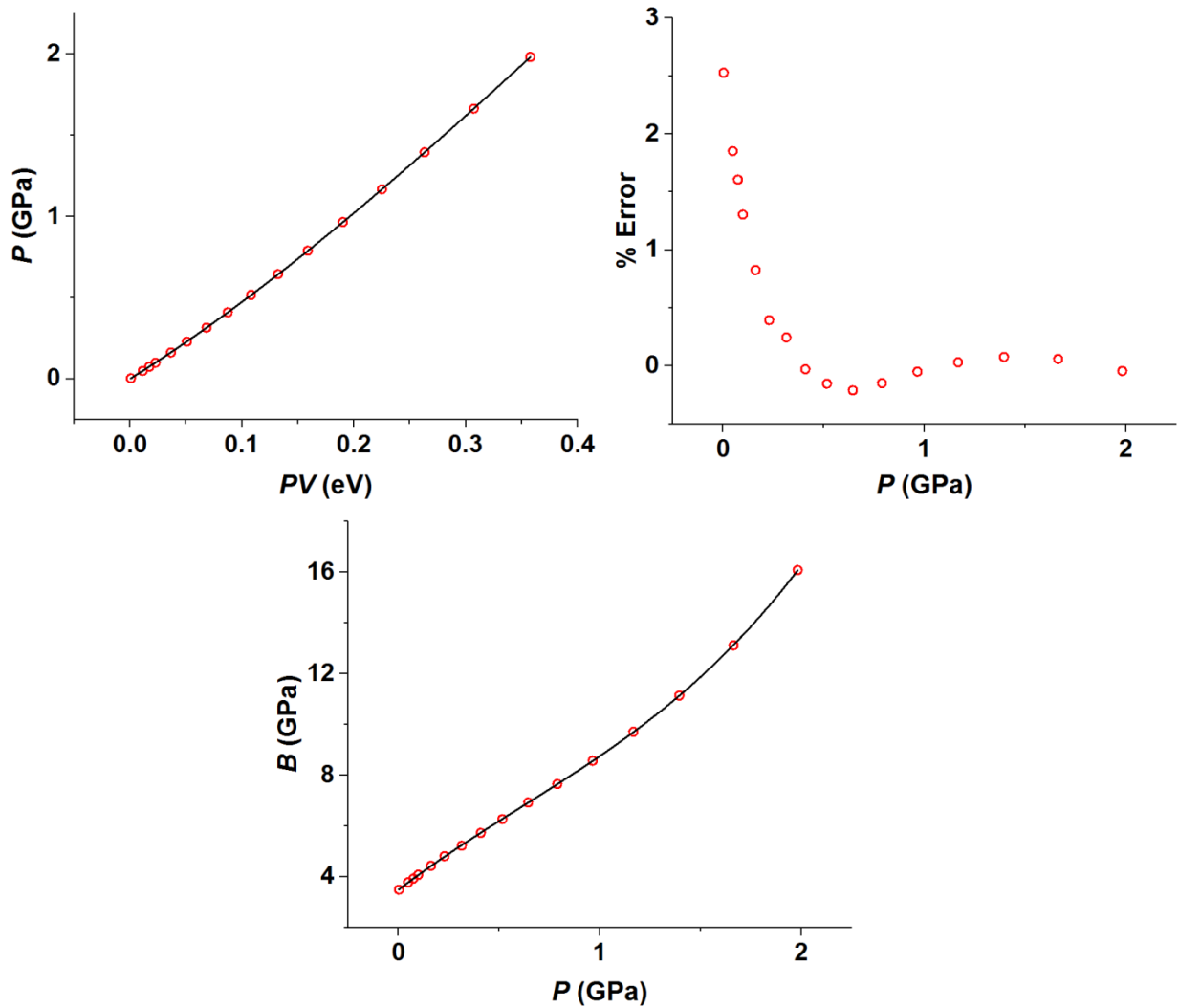

**(j) Kr**

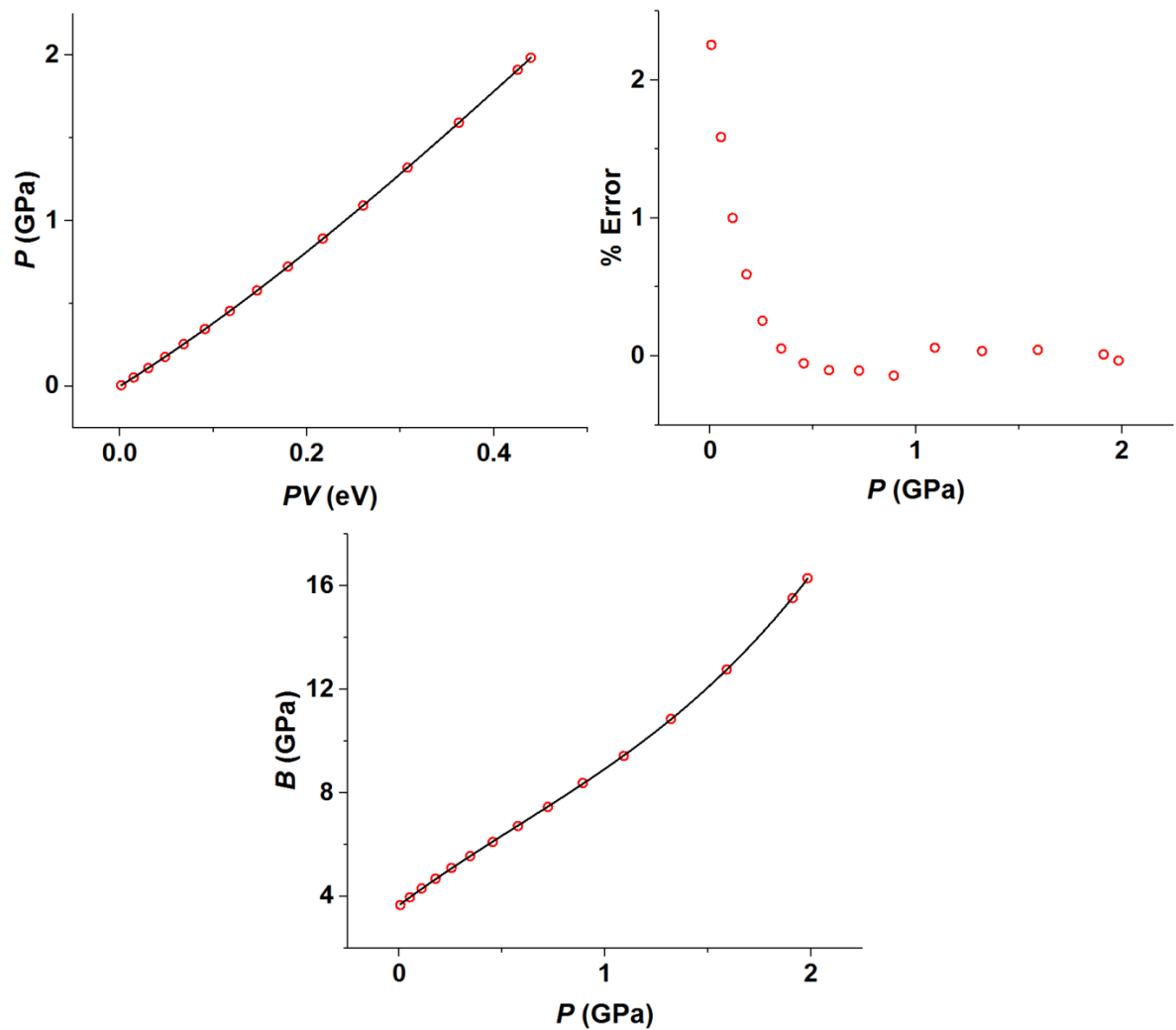

**(k) Xe**

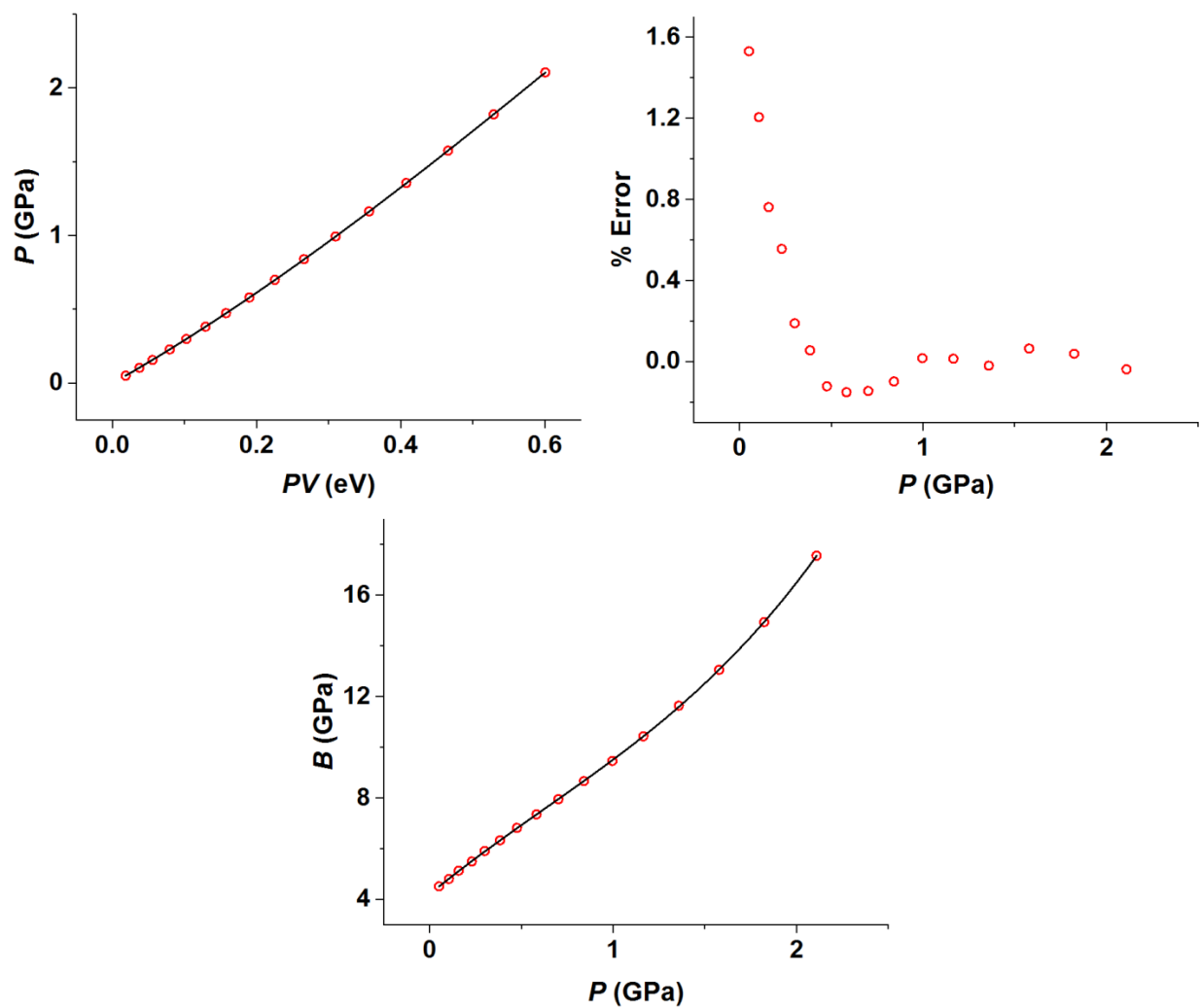

References
(n) Anderson, M. S., & Swenson, C. A., Experimental equations of state for the rare gas solids. *J. Phys. Chem. Solids* **36**, 145-162 (1975).

**(k) H$_2$**

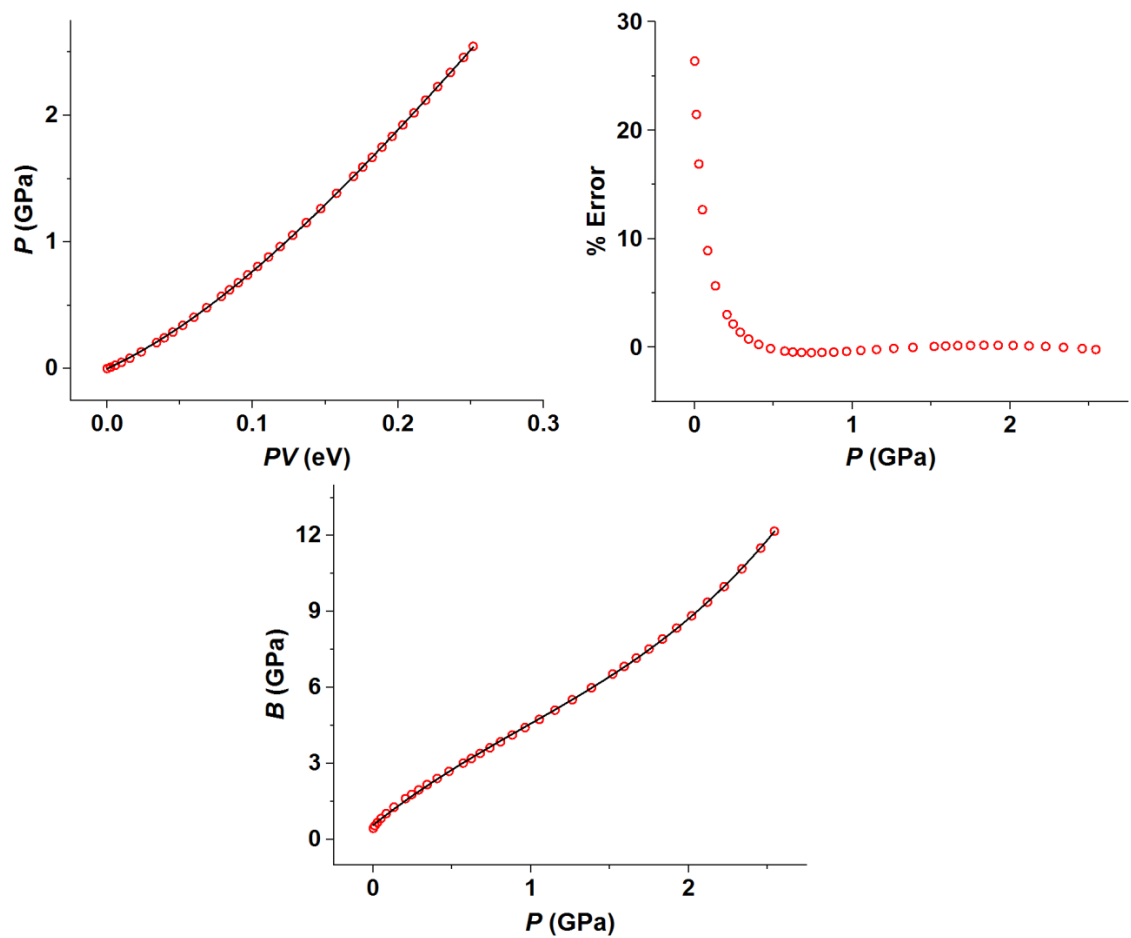

**(l) D$_2$**

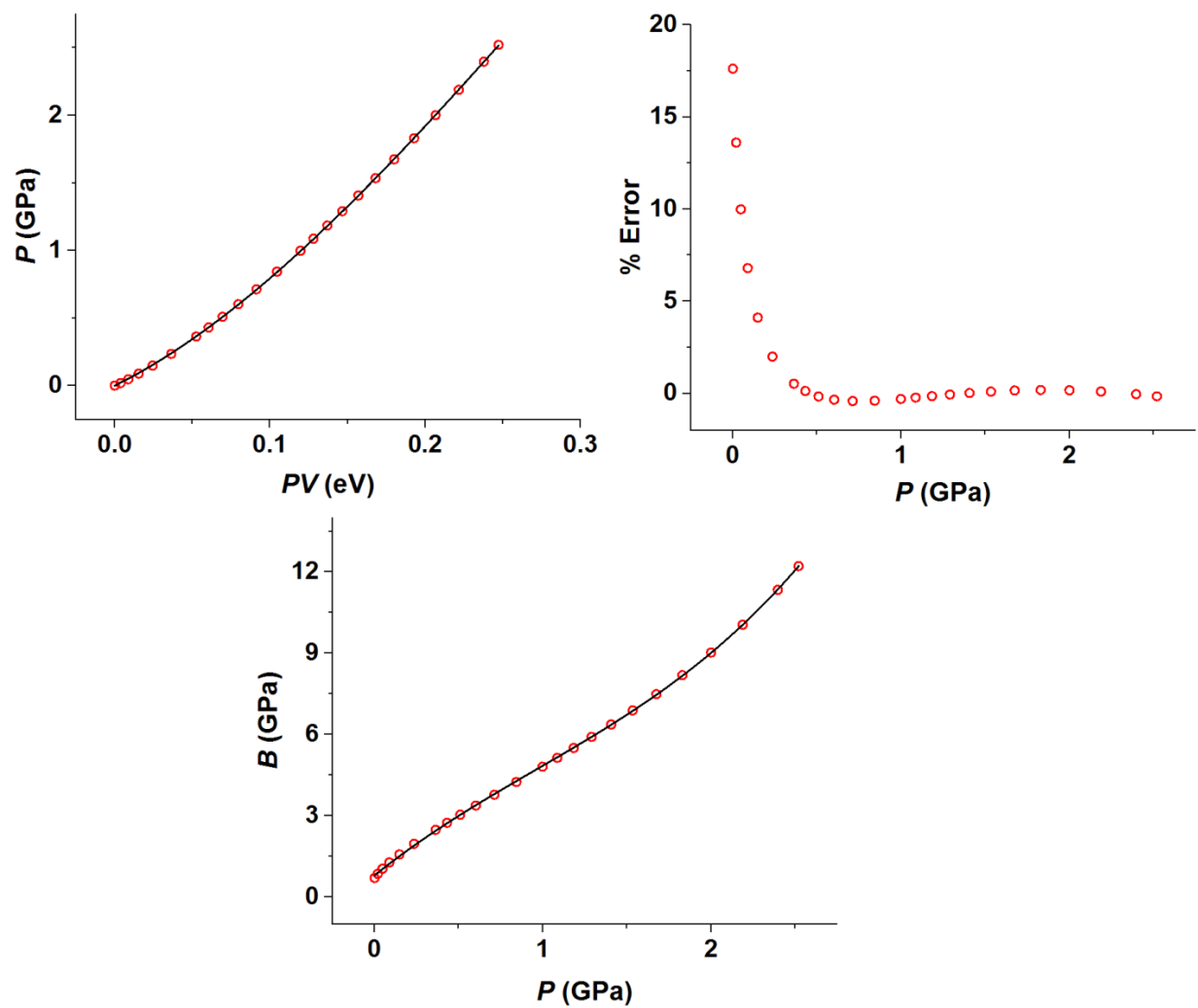

**(m) Polymer PCL**

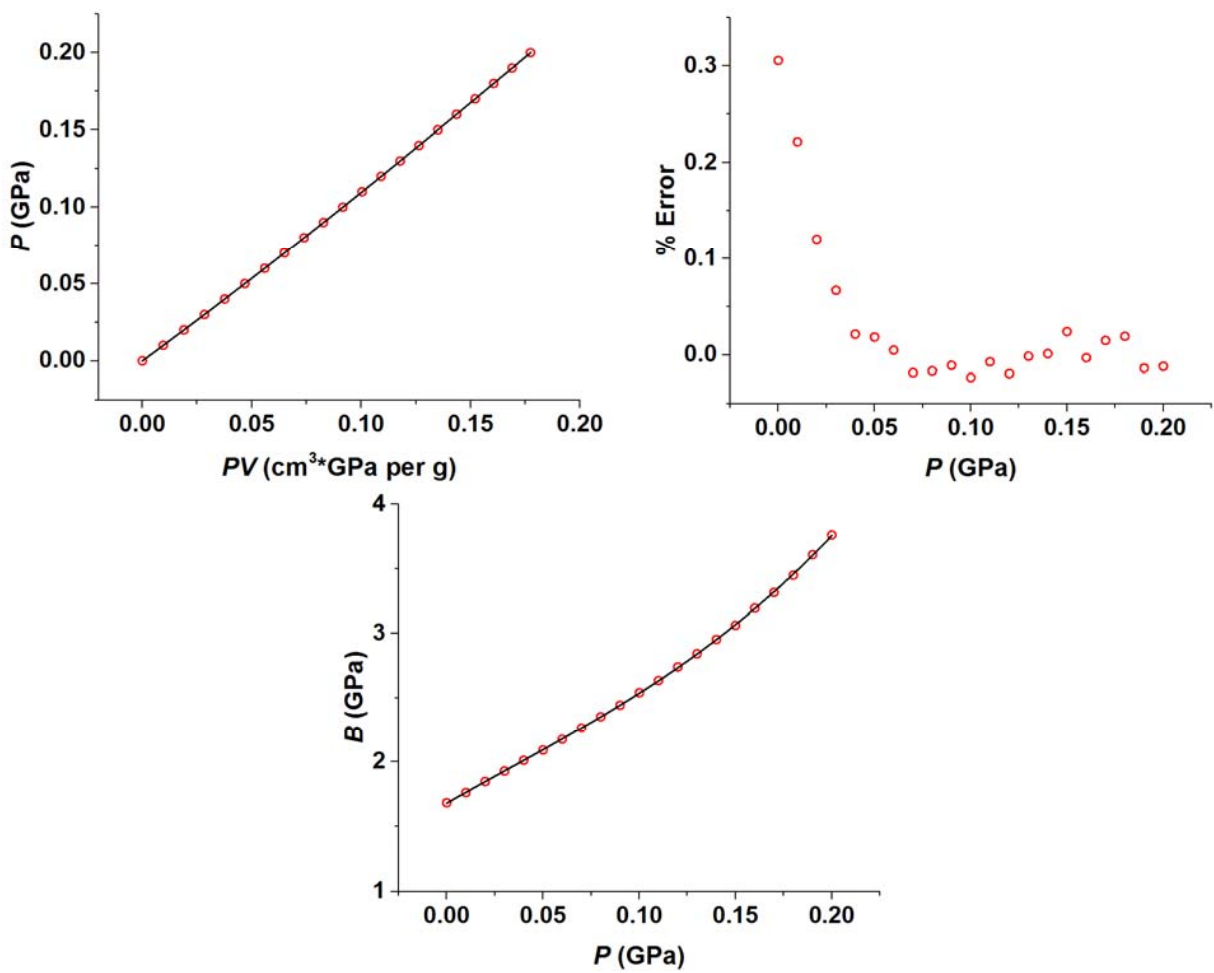

**(n) Liquid H$_2$O at 15 °C**

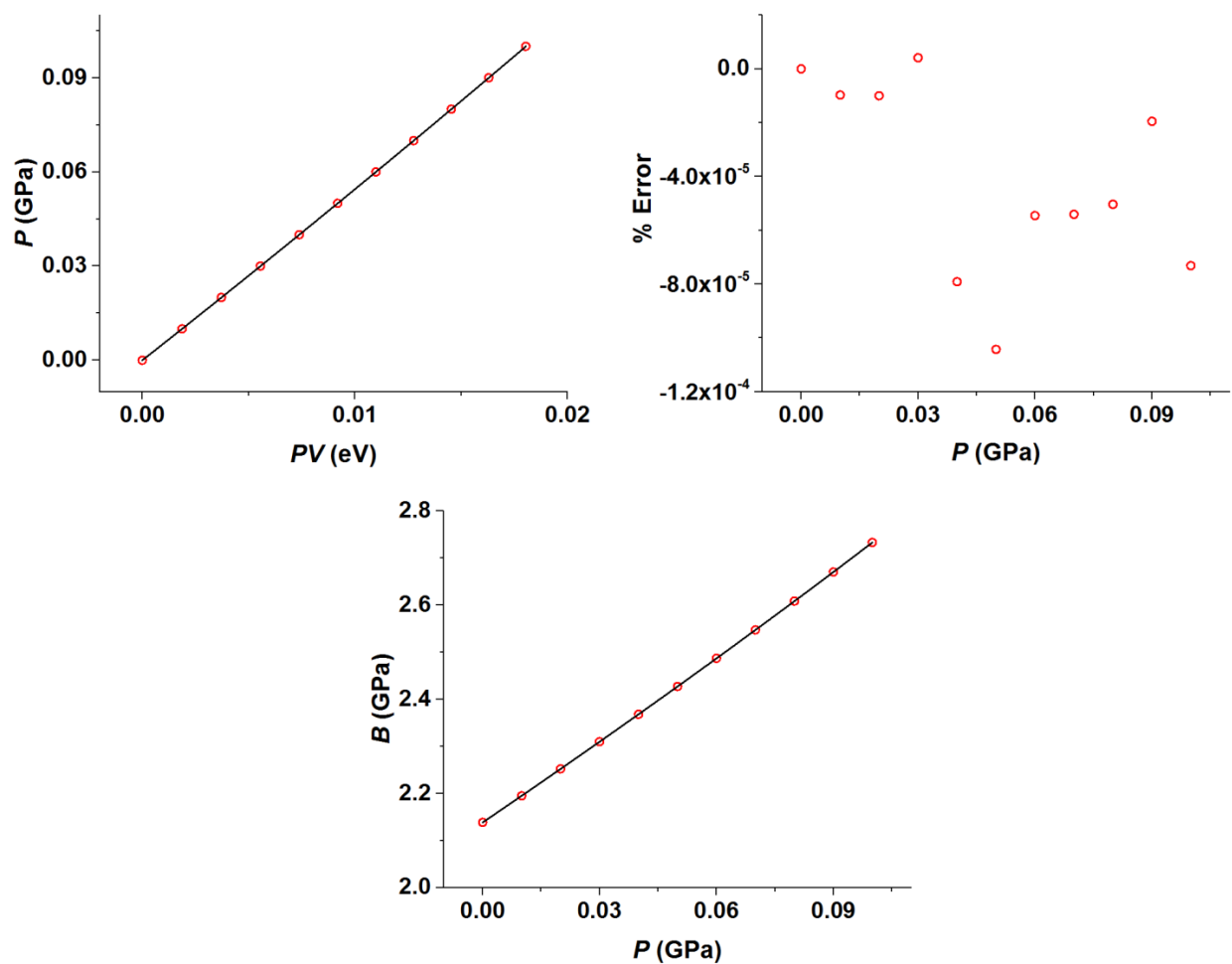

**(o) Liquid $H_2O$ at 25 °C**

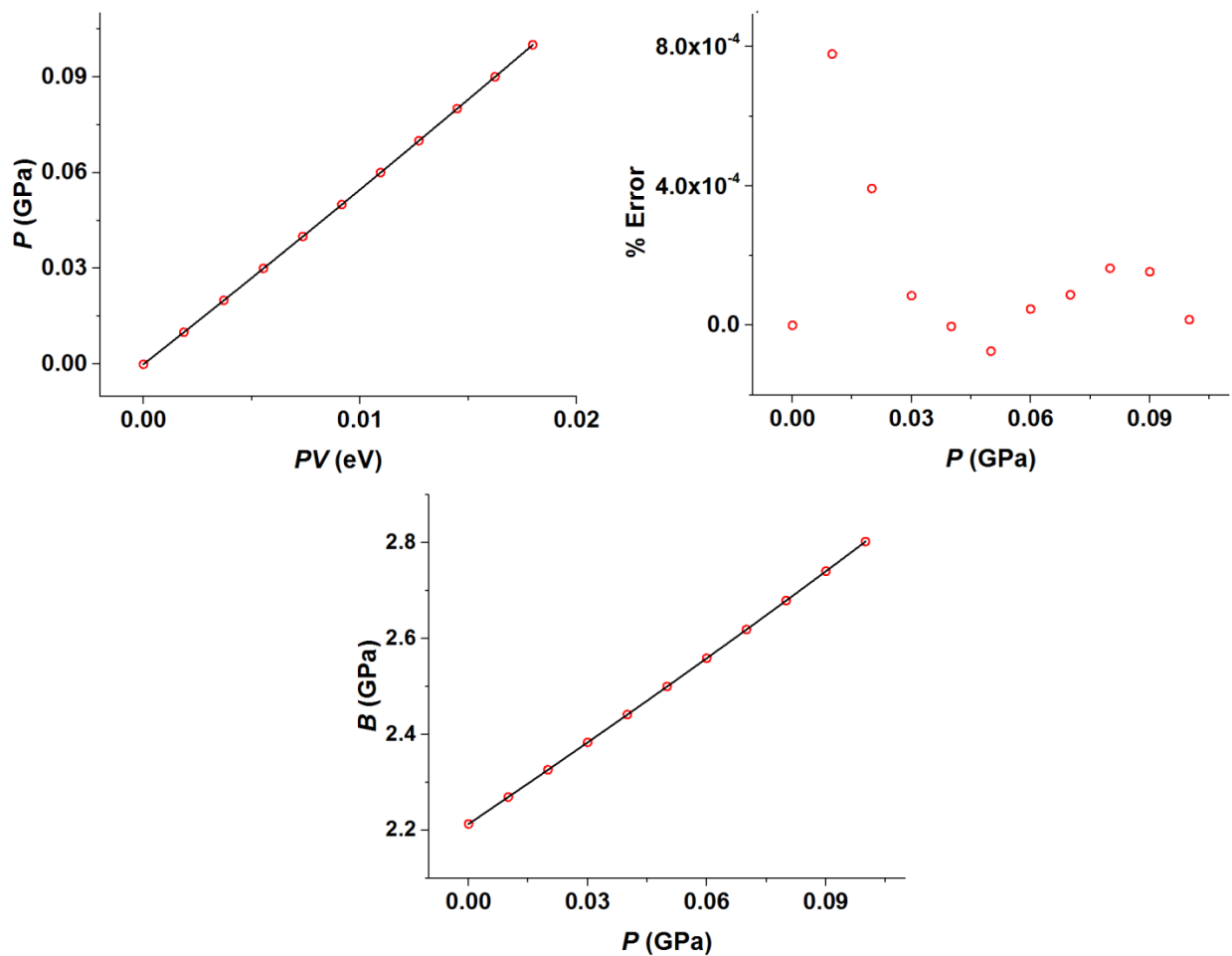

**(p) Liquid H₂O at 35 °C**

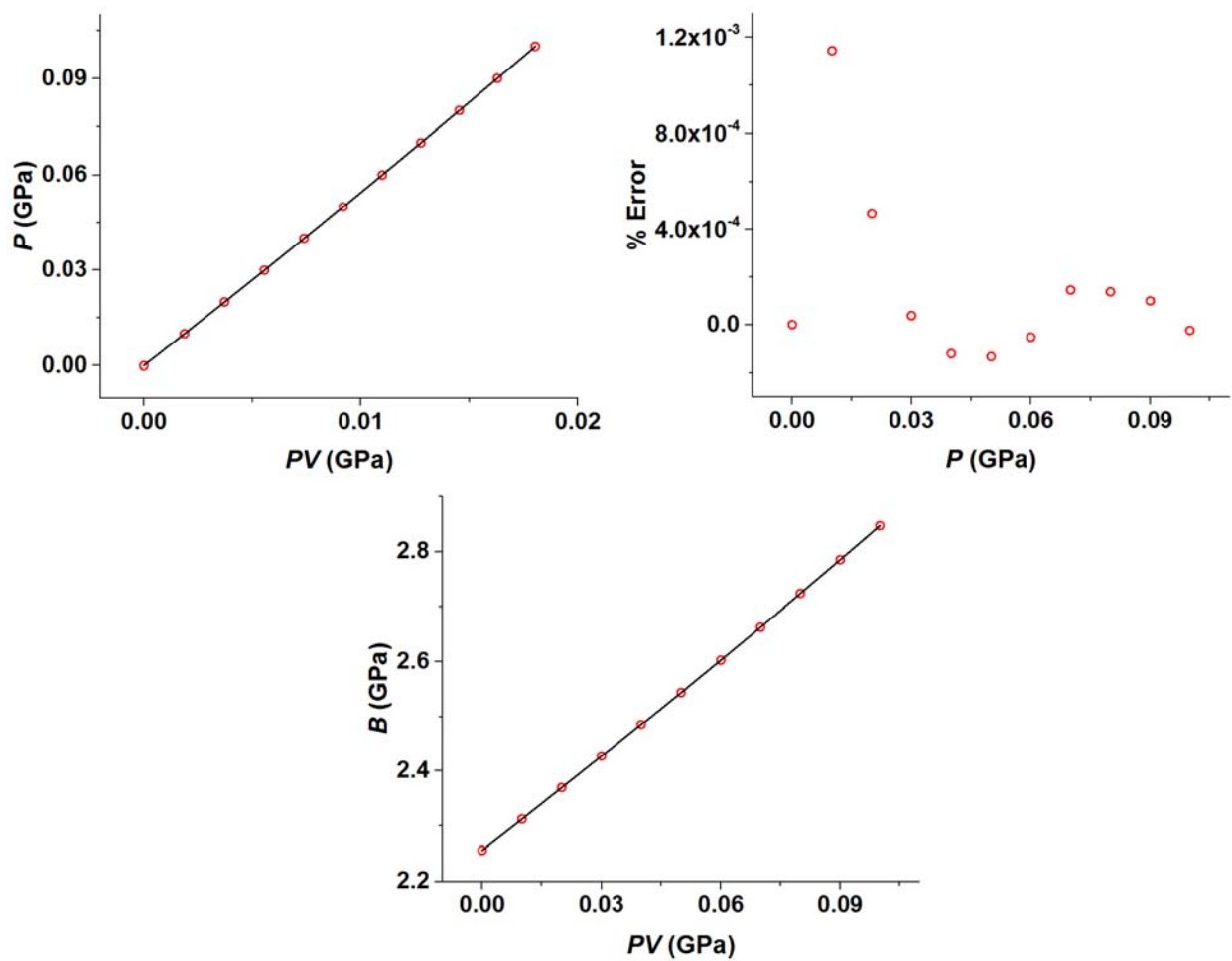

**(q)** *B*(*P*)-vs-*P* plot for MgO

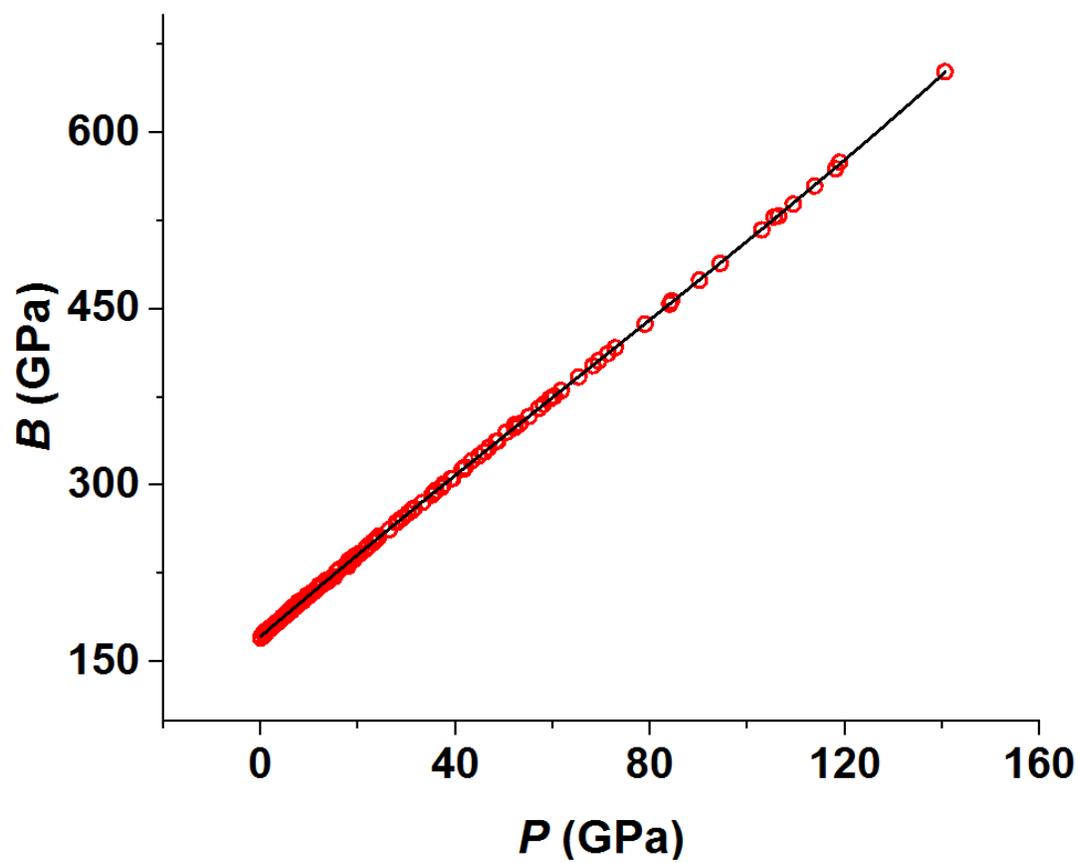

**(r)** *B(P)*-vs-*P* plot for MgSiO$_3$

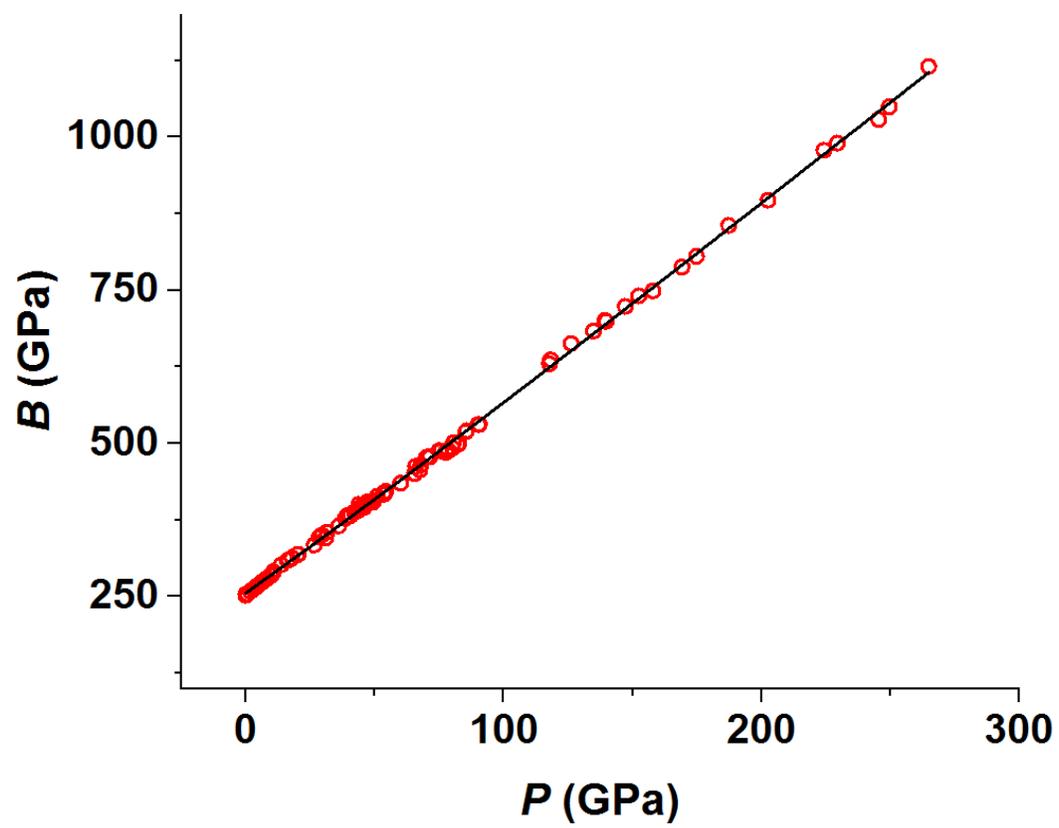

## 4. Analyses of the $B_{0,expt}$ values reported for the various phases of Te, Se and S

For each phase of chalcogen (Te, Se or S) found in a certain pressure region $P_1 - P_2$, the analysis of the *P*-vs-*V* data for the $P_1 - P_2$ region by using the traditional EOS gives rise to the bulk modulus $B_{0,expt}$ and the virtual volume $V_0$ at $P = 0$ even when $P_1$ is well above 0. Namely, $V_0$ is the hypothetical volume the system would have at $P = 0$ if it were to keep the phase found for the $P_1 - P_2$ region down to P = 0. The $B_{0,expt}$ and $V_0$ values found for various phases of Te, Se and S are listed in Table 1 given below.

Table 1. Simulation of the $B_{0,expt}$ values found for the various phases of Te, Se and S in terms our EOS by using their $V_0$ values.

|    | Phase        | $B_{0,expt}$ (GPa) | $V_0$ (Å$^3$) | $B_{0,calc}$ (GPa) |
|----|--------------|--------------------|---------------|--------------------|
|    | Phase I      | 24[14]             | 31.3[14]      | 40.6               |
|    | Phase II     | 54[14]             | 28.0[14]      | 58.4               |
| Te | Phas III     | 57[14]             | 26.2[14]      | 67.5               |
|    | Phase IV     | 115[14]            | 23.6[14]      | 105.9              |
|    | Phase V      | 425[14]            | 20.7[14]      | 604.3              |
|    | Phase I      | 48.1[18]           | 21.6[18]      | 68.6               |
| Se | Phase II     | 63.7[18]           | 14.5[18]      | 60.5               |
|    | Phase III    | 263[18]            | 12.9[18]      | 249.8              |
|    | Phase IV     | 458[18]            | 12.1[18]      | 378.2              |
|    | Orthorhombic | 14.5[21]           | 25.6[21]      | 20.4               |
| S  | BCO          | 21.9[21]           | 23.9[21]      | 28.8               |
|    | β-Po         | 30.6[21]           | 17.2[21]      | 78.0               |

In our EQS analysis of the *P*-vs-*V* data for a system undergoing several phase transitions, we obtain a single *B(P)*-vs-*P* relationship (Eq. 3b) valid for the entire pressure region covering all the phases studied. As already mentioned in the text, Eq. 3b provides only one $B_0$. In order to simulate the $B_{0,\text{expt}}$ values found for the various phases, we proceed as follows:

1) For each phase found in the $P_1 - P_2$ pressure region, we analyze the EOS analysis by using only the $P$-vs-$V$ data of the $P_1 - P_2$ pressure region. Since the pressure region covering each phase is rather narrow, we employ the quadratic approximation of our EOS, namely, $P = \alpha_1(PV) + \alpha_2(PV)^2$.

2) Using the $P = \alpha_1(PV) + \alpha_2(PV)^2$, we determine the $B(V)$-vs-$V$ relationship using Eq. 3a with $\alpha_3 = 0$.

3) Then we evaluate the $B(V)$ value at $V = V_0$. The resulting $B(V_0)$ is now referred to as the calculated $B_{0,\text{calc}}$.

The $B_{0,\text{calc}}$ values calculated as described above are listed in Table 1, which exhibits a reasonable agreement with the $B_{0,\text{calc}}$ and $B_{0,\text{expt}}$ values.